\documentclass[twocolumn,amsmath,amssymb,pre,nobalancelastpage]{revtex4}

\usepackage{amsthm}
\usepackage{newlfont}
\usepackage{graphicx}
\usepackage{subfigure}
\usepackage[section]{placeins}
\usepackage{psfrag}
\usepackage{caption2}
\usepackage{flafter}
\usepackage{epsfig}
\usepackage{epsfig}
\usepackage{bm}%
\def\fnum@figure{\figurename\thefigure}
\renewcommand{\figurename}{Fig.}

\def\@makecaption#1#2{\vskip\abovecaptionskip
  \sbox\@tempboxa{\small #1: #2}%
  \ifdim \wd\@tempboxa >\hsize \small #1: #2\par
  \else \global \@minipagefalse \hb@xt@\hsize{\hfil\box\@tempboxa\hfil}\fi
  \vskip\belowcaptionskip}
\newcommand{\cleqn}{\setcounter{equation}{0}}

\newtheorem{theorem}{Theorem}[section]

\newcommand{\T}{\mathbb{T}}

\makeatletter
    
    \newcommand{\Rmnum}[1]{\expandafter\@slowromancap\romannumeral #1@}
  \makeatother

\def\({\left(}
\def\){\right)}
\def\[{\begin{eqnarray}}
\def\]{\end{eqnarray}}

\def\ep{\epsilon}
\def\La{\Lambda}
\def\la{\lambda}

\def\om{\omega}
\def\De{\Delta}
\begin{document}


\title{Rogue waves of the Hirota and the Maxwell-Bloch equations}
\author{Chuanzhong Li$^{a}$}
\author{Jingsong He$^{a}$}
\email{hejingsong@nbu.edu.cn}
\author{K. Porseizan$^{b}$}

\address{$^{a}$Department of
Mathematics,  Ningbo University, Ningbo, 315211, China \\$^{b}$Department of Physics, Pondicherry University, Puducherry 605014, India.}

\begin{abstract}
In this paper, we derive a Darboux transformation of the Hirota and the Maxwell-Bloch(H-MB) system which is governed by femtosecond pulse propagation through an erbium doped fibre and further generalize it to the matrix form of the $n$-fold Darboux transformation of this system. This  $n$-fold Darboux transformation implies the determinant representation of $n$-th new solutions of $(E^{[n]},p^{[n]}, \eta^{[n]})$ generated from known solution of $(E, p,\eta )$. The determinant representation of
$(E^{[n]},p^{[n]} ,\eta^{[n]})$ provides soliton solutions, positon solutions, and  breather solutions (both bright and dark breathers) of the H-MB system. From the breather solutions, we also construct bright and dark rogue wave solutions for the H-MB system, which is currently one of the hottest topics in mathematics and physics. Surprisingly, the rogue wave solution for $p\, and\, \eta$ has two peaks because of the order of the numerator and denominator of them. Meanwhile, after fixing time and spatial parameters and changing other two unknown parameters $\alpha$ and $\beta$, we generate a rogue wave shape for the first time.

\end{abstract}

\maketitle

\section{Introduction}

In the past four decades, nonlinear science has experienced an explosive growth with the invention of several exciting and fascinating new concepts such as solitons, dromions, positons, rogue waves, similaritons, supercontinuum generation, complete integrability, fractals, chaos etc. Many of the completely integrable nonlinear partial differential systems (NPDEs) admit one of the most striking aspects of nonlinear phenomena, described as the soliton, a universal character and is of great mathematical interest. The study of the solitons and other related solutions like positons have become one of the most exciting and extremely active areas of research in the field of nonlinear sciences.

Among all concepts, in addition to solitons and positons \cite {Matveev92pla,Matveev92pla2,Matveev02TMP,inhomogeneousHMB},  rogue waves have also been not only the subject of intensive research in oceanography \cite{C.Kharif,Akhmediev,Osborne} but also they have been studied extensively in several other areas, such as matter rogue wave \cite{MR,W.M.Liu} in Bose-Einstein condensates, rogue waves in surface and space plasmas \cite{PR}, financial rogue waves describing the possible physical mechanisms in financial markets and related fields \cite{FR}. In some of the above fields, soliton system such as nonlinear Schr\"odinger (NLS) equation \cite{Peregrine}, derivative NLS system \cite{shuweiJPA,shuweiJMP} and so on are considered and reported to admit rogue wave solutions under a certain specific choice of parameters.  It has been proved that modulational instability is one of the main generating mechanisms for the rogue waves \cite{Peregrine,shuweiJPA,shuweiJMP,Kristian,Zakharov,Zakharov2} and can be well-described by the analytical expressions for the spectra of breather solutions at the point of extreme compression.

In 1967, McCall and Hahn \cite{McCallPRL} explored a special type of lossless pulse propagation in two-level resonant media. They have discovered the self-induced transparency (SIT) effect which can be explained by  using the Maxwell-Bloch (MB) system.   If we consider these effects in erbium doped nonlinear fibre, the system will be governed by the coupled system of the NLS and the MB equation (NLS-MB system)\cite{Nakazawa,Nakazawa2,NLSMBPorsezian,heTheor,hexuNLSMB}.

Rogue waves have been reported in different branches of physics, where the system dynamics is governed mostly by a single nonlinear partial differential equation \cite{shuweiJPA,shuweiJMP}. But our main interest is to analyze the possibility of rogue waves in coupled nonlinear systems. The higher-order NLS and Maxwell-Bloch (HNLS-MB) system as a higher-order correction of NLS-MB system were  shown to admit Lax pair and  soliton-type pulse propagation \cite{Nakkeeran95,Nakkeeran95jmodopt,Nakkeeranjpa}. Kodama \cite{Kodama} has shown that with a suitable transformation, higher order NLS equation  can be reduced to the Hirota equation \cite{hirotaeq} whose rogue wave solution has already been reported in \cite{Akhmedievhirota,taohirota}. In a similar way, after suitable choice of self-steepening and self-frequency effects, we obtain the H-MB system in the following form \cite{PorsezianPRL}:

\begin{eqnarray}\notag
E_z&=&i\alpha(\frac12E_{tt}+|E|^2E)+\beta(E_{ttt}+6|E|^2E_t)+2p,\\
\\
p_t&=&2i\omega p+2E \eta,\\
\eta_t&=&-(Ep^*+E^*p),
\end{eqnarray}
where $E$ is the normalized slowly varying amplitude of the complex field envelope, $p$ is the polarization, $\eta$ means the population inversion, $(\omega, \alpha, \beta)$ are three real constants and $*$ represents complex conjugate. $\beta$ represents the strength of the higher order linear and nonlinear effects.

The H-MB system has been shown to be integrable and also admits a Lax pair and other required properties
of complete integrability \cite{PorsezianPRL}.
 Among many analytical methods, it is well known that the Darboux
 transformation is one of the efficient methods to generate the
 soliton solutions for integrable systems \cite{Matveev}.
 The determinant representation of $n$-fold Darboux transformation
 of the Ablowitz-Kaup-Newell-Segur (AKNS) system was given in \cite{Hedeterminant,hehrwnls}.
 The main task of this paper will be to construct $n$-fold Darboux
 transformation of the H-MB system and find different kinds of
 solutions of the H-MB system using the Darboux transformation.

The paper is organized as follows.  In section 2, the Lax representation of H-MB system is introduced.   In section 3, we derived the one-fold Darboux transformation of the H-MB system. In section 4, the generalization of one-fold Darboux transformation to $n$-fold Darboux transformation of the H-MB system will be given.  Using these Darboux transformations, one soliton, two soliton and positon solutions are derived in section 5 $\&$ 6 by assuming trivial seed solutions. In section 7, starting from a periodic seed solution, breather solution of the H-MB system is provided. A Taylor expansion from breather solution will help us to construct the rogue wave solution in section 8. Section 9 is devoted to conclusion and discussions.

\section{ Lax representation of the H-MB system  }\cleqn
 In this section, we will concentrate on the linear eigenvalue problem of the Hirota and the Maxwell-Bloch(H-MB) system. The linear eigenvalue problem is expressed in the form of the Lax pair $U$ and $V$ as

\begin{eqnarray}\label{linear}
\Phi_t&=&U\Phi,\ \ \ \ \
\Phi_z=V\Phi,
\end{eqnarray}

where
\begin{widetext}
\begin{eqnarray}
U&=&\lambda\left(\begin{matrix}-i& 0\\ 0& i
\end{matrix}\right)+\left(\begin{matrix}0& E\\ -E^*& 0
\end{matrix}\right)=-i\lambda \sigma_3+U_0,\ \ \ \sigma_3=\left(\begin{matrix}1& 0\\ 0& -1
\end{matrix}\right),\\
V&=&\lambda^3\left(\begin{matrix}4i\beta& 0\\ 0& -4i\beta
\end{matrix}\right)+\lambda^2\left(\begin{matrix}-\alpha i& -4\beta E\\ 4\beta E^*& \alpha i
\end{matrix}\right)+\lambda\left(\begin{matrix}-2\beta i|E|^2 &\alpha E-2\beta iE_t\\ -\alpha E^*-2\beta iE_t^*& 2\beta i|E|^2
\end{matrix}\right)\\ \notag
&&+\left(\begin{matrix}\frac{\alpha}2i|E|^2-\beta (EE^*_t-E_tE^*) &2\beta |E|^2E+\frac{\alpha}2iE_t+\beta E_{tt}\\ -2\beta|E|^2E^*+\frac{\alpha}2iE^*_t-\beta E^*_{tt}& -\frac{\alpha}2i|E|^2+\beta (EE^*_t-E_tE^*)
\end{matrix}\right)+i\frac{1}{\lambda+\omega}\left(\begin{matrix}\eta & -p\\ -p^*& -\eta
\end{matrix}\right)\\
&:=&\lambda^3V_3+\lambda^2V_2+\lambda V_1+V_0+i\frac{1}{\lambda+\omega}V_{-1},\\
\Phi &= &\Phi(\lambda)=(\begin{matrix}
\Phi_1(\lambda,t,z)\\
\Phi_2(\lambda,t,z)
\end{matrix})
\end{eqnarray}
\end{widetext}
 is an eigenfunction associated with eigenvalue parameter $\lambda$ of the linear Eq. (\ref{linear}),  and $V_i$ denotes the coefficient matrix of term
$\lambda^i$. We obtain the classical Hirota and the Maxwell-Bloch system when $\alpha=2, \beta=-1$. Being different from AKNS system, only a part of the $V$ matrix is polynomials in terms of $E$ and its $t$ derivatives in this system. Using the above linear system of the H-MB system, one-fold Darboux transformation will be introduced in the next section.

\section{ One-fold Darboux transformation for the H-MB system  }\cleqn

In this section, we construct and prove the one-fold Darboux transformation for the H-MB system. First, we consider the transformation about linear function $\Phi$ in the form
\begin{eqnarray}
\Phi'&=&T\Phi=(\lambda A-S)\Phi,
\end{eqnarray}
where
\begin{eqnarray}
A=\left(\begin{matrix}a_{11}& a_{12}\\ a_{21}& a_{22}
\end{matrix}\right),\ \ \
S=\left(\begin{matrix}s_{11}& s_{12}\\ s_{21}& s_{22}
\end{matrix}\right).
\end{eqnarray}

The new function $\Phi'$ satisfies
\begin{eqnarray}
\Phi'_t&=&U'\Phi',\\
\Phi'_z&=&V'\Phi'.
\end{eqnarray}

Then the matrix $T$ should satisfy the following identities
\begin{eqnarray}\label{tequation}
T_t+TU&=&U'T,\\ \label{zequation}
T_z+TV&=&V'T.
\end{eqnarray}
Substituting the matrices $A$ and $S$ into Eq. \eqref{tequation} and comparing the coefficients of both sides will lead to the following conditions
\[a_{12}=a_{21}=0,\ \ \ (a_{11})_t=(a_{22})_t=0.\]

For our further discussions, we choose $A=I$ and $T=(\lambda I-S)$. The relation between old solutions $(E, p,\eta)$ and new solutions $(E', p',\eta')$, which is called Darboux transformation, can be obtained by using  Eqs. \eqref{tequation} and  \eqref{zequation}.

From Eq. \eqref{tequation}, we have
\small
\[\label{E' and E}
E'&=&E-2is_{12},\ \ \ s_{21}=-s^*_{12}\\
\notag S_t&=&\left(\begin{matrix}0& E\\ -E^*& 0
\end{matrix}\right)S-S\left(\begin{matrix}0& E\\ -E^*& 0
\end{matrix}\right)+i[S,\sigma_3]S.\\
\]
\normalsize
Similarly, using Eq. \eqref{zequation}, we obtain the following set of relations
\begin{eqnarray}
\notag
&&-S_z+(\lambda I-S)(\lambda^3V_3+\lambda^2V_2+\lambda V_1+V_0+\frac{iV_{-1}}{\lambda+\om})\\
\notag
&&=(\lambda^3V_3+\lambda^2V'_2+\lambda V'_1+V'_0+\frac{iV'_{-1}}{\lambda+\om})(\lambda I-S).\\\label{zeqdetailed}
\end{eqnarray}
Multiplying both sides of Eq.\eqref{zeqdetailed} by $\la I-S$ will lead to

\begin{eqnarray*}
&&-S_z(\la+\om)+(\la I-S)(-4i\la^3(\la+\om)\sigma_3\\
&&+\la^2(\la+\om)V_2+\la(\la+\om) V_1+V_0(\la+\om)+iV_{-1})\\
&=&(-4i\la^3(\la+\om)\sigma_3+\la^2(\la+\om)V'_2\\
&&+\la (\la+\om) V'_1+V'_0(\la+\om)+iV'_{-1})(\la I-S).
\end{eqnarray*}

Collecting the different powers of $\la$,  we obtain the following set of identities\\
$\la^0$:
\begin{eqnarray}\notag
S_z
&=&( V'_0+i\om^{-1}V'_{-1})S-S(i\om^{-1}V_{-1}+V_0),\\ \label{la0}
\end{eqnarray}
$\la$:
\begin{eqnarray}\notag
S_z
&=&(\om V_0+iV_{-1})
-S(\om V_1+V_0)+(\om V'_1+V'_0)S\\ \label{la1}
 &&+(-\om V'_0-iV'_{-1}).
\end{eqnarray}
$\la^2$:
\begin{eqnarray}\label{la2}
&&(\om V_1+V_0)-S(\om V_2+V_1)\\ \notag
&=&(\om V'_1+V'_0)-(\om V'_2+V'_1)S.
\end{eqnarray}
$\la^3$:
\begin{eqnarray}\label{la3}
&&(\om V_2+V_1)-S(\om V_3+V_2)\\ \notag
&=&(\om V'_2+V'_1)-(\om V_3+V'_2)S,
\end{eqnarray}
$\la^4$:
\begin{eqnarray}\label{la4}
V'_2=V_2-[S,V_3].
\end{eqnarray}

From the above identities, after simplifications, we get
\[\label{E'andE2}
E'&=&E-2 is_{12},
\]
\begin{eqnarray}\label{V'-1dressing}
V'_{-1}=(S+\om) V_{-1}(S+\om)^{-1},
\end{eqnarray}
which gives one-fold Darboux transformation of the H-MB system later.

We suppose
\[\label{SandH}
S=H\La H^{-1}\]
where
$\La=\left(\begin{matrix}\la_1& 0\\ 0& \la_2
\end{matrix}\right)$,\\
 $H=\left(\begin{matrix}\Phi_1(\la_1,t,z)& \Phi_1(\la_2,t,z)\\ \Phi_2(\la_1,t,z)& \Phi_2(\la_2,t,z)
\end{matrix}\right):=\left(\begin{matrix}\Phi_{1,1}& \Phi_{1,2}\\ \Phi_{2,1}& \Phi_{2,2}
\end{matrix}\right).$

In order to satisfy the constraints of $S$ and $V'_{-1}$ which is similar to $V_{-1}$, i.e. $s_{21}=-s^*_{12},$ following constraints will be used
\[\label{constraint-lam}
\la_2&=&\la_1^*,\ \ s_{11}=s^*_{22},\\\label{H}
H&=&\left(\begin{matrix}\Phi_1(\la_1,t,z)& -\Phi^*_2(\la_1,t,z)\\ \Phi_2(\la_1,t,z)& \Phi^*_1(\la_1,t,z)
\end{matrix}\right).\]

After  tedious calculations, Eqs. \eqref{E'andE2}-\eqref{H} and Eq. \eqref{linear} will lead to Eq. \eqref{tequation} and Eq. \eqref{zequation}, i.e. the transformation Eq. \eqref{E'andE2} and Eq. \eqref{V'-1dressing} with the conditions Eq. \eqref{constraint-lam}, Eq. \eqref{H} is the Darboux transformation of the H-MB system.

The detailed form of one-fold Darboux transformation of the H-MB system in terms of eigenfunctions will be given in the next section.

\section{Determinant representation of $n$-fold Darboux transformation}\cleqn

In this section, we will construct the determinant representation of the $n$-fold Darboux transformation of the H-MB system. For this purpose, we introduce n eigenfunctions
\[\left(\begin{matrix}\Phi_{1,i}\\
 \Phi_{2,i}\end{matrix}\right)=\Phi|_{\la=\la_i},\, i=1,2,\dots,2n\]
with  the following constraint on the eigenvalues $\la_{2n-1}=\la_{2n}^*$ and  the reduction conditions on eigenfunctions as $\Phi_{2,2n}=\Phi_{1,2n-1}^* and \ \ \Phi_{2,2n-1}=-\Phi_{1,2n}^*$. For our further discussions, this reduction condition has been used.

For completeness, as the simplest Darboux transformation, the determinant representation of one-fold Darboux transformation of the H-MB system will be introduced in the following theorem using identities \eqref{E'andE2} and  \eqref{V'-1dressing}.

The one-fold Darboux transformation of the H-MB system is expressed as

\begin{eqnarray}\label{Edetail} E^{[1]}&=&E-2i\frac{(\la_1-\la^*_1)\Phi_{1,1}\Phi_{2,1}^*}{\Delta_1},\\ \notag
p^{[1]}&=&\frac{1}{T_{\De_1}}(2\eta[(-\om-\la_1)|\Phi_{1,1}|^2+(-\om-\la_1^*)|\Phi_{2,1}|^2]\\ \notag
&&(\la_1^*-\la_1)\Phi_{1,1}\Phi_{2,1}^*-p^*(\la_1^*-\la_1)^2\Phi_{1,1}^2\Phi_{2,1}^{*2}\\ \notag
&&+p[(-\om-\la_1)|\Phi_{1,1}|^2+(-\om-\la_1^*)|\Phi_{2,1}|^2]^2),\\\label{pdetail}\\ \notag
\eta^{[1]}&=&\frac{1}{T_{\De_1}}[\eta([(\om+\la_1)|\Phi_{1,1}|^2+(\om+\la_1^*)|\Phi_{2,1}|^2]\\ \notag
&&[(\om+\la_1^*)|\Phi_{1,1}|^2+(\om+\la_1)|\Phi_{2,1}|^2]\\
\notag
&&+(\la_1^*-\la_1)^2|\Phi_{1,1}|^2|\Phi_{2,1}|^2)\\ \notag
&&
-p^*(\la_1-\la_1^*)\Phi_{1,1}\Phi_{2,1}^*[(\om+\la_1^*)|\Phi_{1,1}|^2\\
\notag&&+(\om+\la_1)|\Phi_{2,1}|^2]+p[(\om+\la_1)|\Phi_{1,1}|^2\\ \label{etadetail}
&&+(\om+\la_1^*)|\Phi_{2,1}|^2](\la_1-\la_1^*)\Phi_{1,1}^*\Phi_{2,1}],
\end{eqnarray}

where

\begin{eqnarray}\notag
T_{\De_1}&=&[(\om+\la_1)|\Phi_{1,1}|^2+(\om+\la_1^*)|\Phi_{2,1}|^2]\\ \notag
&&[(\om+\la_1^*)|\Phi_{1,1}|^2+(\om+\la_1)|\Phi_{2,1}|^2]\\
&&-(\la_1^*-\la_1)^2|\Phi_{1,1}|^2|\Phi_{2,1}|^2.
\end{eqnarray}

It can be easily proved that the new solution $\eta^{[1]}$ is always real. This one-fold transformation will be used to generate the one-soliton solution from trivial seed solutions of the H-MB system. Also this one-fold Darboux transformation can be further generalized to construct the $n$-fold Darboux transformation of the H-MB system which is proposed in the following theorem.
\begin{theorem}
The $n$-fold Darboux transformation of the H-MB system can be represented as

\begin{eqnarray}
&&T_n(\la;\la_1,\la_2,\la_3,\la_4,\dots,\la_{2n})\\
&=&\la^n I+t_{n-1}^{[n]}\la^{n-1}+\dots+t_1^{[n]}\la+t_0^{[n]}\\
&=&\frac{1}{\De_n}\left(\begin{matrix}(\T_n)_{11}&(\T_n)_{12}\\
(\T_n)_{21}&(\T_n)_{22}\,\,\,\,\,
\end{matrix}\right)
\end{eqnarray}
where
\begin{widetext}
\[ \notag \De_n&=&
\left|\begin{matrix}\Phi_{1,1}&\Phi_{2,1}&\la_1\Phi_{1,1}&\la_1\Phi_{2,1}&\dots&\la_1^{n-1}\Phi_{1,1}&\la_1^{n-1}\Phi_{2,1}\\
\Phi_{1,2}&\Phi_{2,2}&\la_2\Phi_{1,2}&\la_2\Phi_{2,2}&\dots&\la_2^{n-1}\Phi_{1,2}&\la_2^{n-1}\Phi_{2,2}\\
\Phi_{1,3}&\Phi_{2,3}&\la_3\Phi_{1,3}&\la_3\Phi_{2,3}&\dots&\la_3^{n-1}\Phi_{1,3}&\la_3^{n-1}\Phi_{2,3}\\
\Phi_{1,4}&\Phi_{2,4}&\la_4\Phi_{1,4}&\la_4\Phi_{2,4}&\dots&\la_4^{n-1}\Phi_{1,4}&\la_4^{n-1}\Phi_{2,4}\\
\vdots&\vdots&\vdots&\vdots&\vdots&\vdots&\vdots\\
\Phi_{1,2n-1}&\Phi_{2,2n-1}&\la_{2n-1}\Phi_{1,2n-1}&\la_{2n-1}\Phi_{2,2n-1}&\dots&\la_{2n-1}^{n-1}\Phi_{1,2n-1}&\la_{2n-1}^{n-1}\Phi_{2,2n-1}\\
\Phi_{1,2n}&\Phi_{2,2n}&\la_{2n}\Phi_{1,2n}&\la_{2n}\Phi_{2,2n}&\dots&\la_{2n}^{n-1}\Phi_{1,2n}&\la_{2n}^{n-1}\Phi_{2,2n}\\
\end{matrix}\right|\]
{\tiny \[\notag(\T_n)_{11}&=&
\left|\begin{matrix}1&0&\la&0&\dots&\la^{n-1}&0&\la^n\\
\Phi_{1,1}&\Phi_{2,1}&\la_1\Phi_{1,1}&\la_1\Phi_{2,1}&\dots&\la_1^{n-1}\Phi_{1,1}&\la_1^{n-1}\Phi_{2,1}&\la_1^{n}\Phi_{1,1}\\
\Phi_{1,2}&\Phi_{2,2}&\la_2\Phi_{1,2}&\la_2\Phi_{2,2}&\dots&\la_2^{n-1}\Phi_{1,2}&\la_2^{n-1}\Phi_{2,2}&\la_2^{n}\Phi_{1,2}\\
\Phi_{1,3}&\Phi_{2,3}&\la_3\Phi_{1,3}&\la_3\Phi_{2,3}&\dots&\la_3^{n-1}\Phi_{1,3}&\la_3^{n-1}\Phi_{2,3}&\la_3^{n}\Phi_{1,3}\\
\Phi_{1,4}&\Phi_{2,4}&\la_4\Phi_{1,4}&\la_4\Phi_{2,4}&\dots&\la_4^{n-1}\Phi_{1,4}&\la_4^{n-1}\Phi_{2,4}&\la_4^{n}\Phi_{1,4}\\
\vdots&\vdots&\vdots&\vdots&\vdots&\vdots&\vdots&\vdots\\
\Phi_{1,2n-1}&\Phi_{2,2n-1}&\la_{2n-1}\Phi_{1,2n-1}&\la_{2n-1}\Phi_{2,2n-1}&\dots&\la_{2n-1}^{n-1}\Phi_{1,2n-1}&\la_{2n-1}^{n-1}\Phi_{2,2n-1}&\la_{2n-1}^{n}\Phi_{1,2n-1}\\
\Phi_{1,2n}&\Phi_{2,2n}&\la_{2n}\Phi_{1,2n}&\la_{2n}\Phi_{2,2n}&\dots&\la_{2n}^{n-1}\Phi_{1,2n}&\la_{2n}^{n-1}\Phi_{2,2n}&\la_{2n}^{n}\Phi_{1,2n}
\end{matrix}\right|\]}
{\tiny \[\notag(\T_n)_{12}&=&
\left|\begin{matrix}0&1&0&\la&\dots&0&\la^{n-1}&0\\
\Phi_{1,1}&\Phi_{2,1}&\la_1\Phi_{1,1}&\la_1\Phi_{2,1}&\dots&\la_1^{n-1}\Phi_{1,1}&\la_1^{n-1}\Phi_{2,1}&\la_1^{n}\Phi_{1,1}\\
\Phi_{1,2}&\Phi_{2,2}&\la_2\Phi_{1,2}&\la_2\Phi_{2,2}&\dots&\la_2^{n-1}\Phi_{1,2}&\la_2^{n-1}\Phi_{2,2}&\la_2^{n}\Phi_{1,2}\\
\Phi_{1,3}&\Phi_{2,3}&\la_3\Phi_{1,3}&\la_3\Phi_{2,3}&\dots&\la_3^{n-1}\Phi_{1,3}&\la_3^{n-1}\Phi_{2,3}&\la_3^{n}\Phi_{1,3}\\
\Phi_{1,4}&\Phi_{2,4}&\la_4\Phi_{1,4}&\la_4\Phi_{2,4}&\dots&\la_4^{n-1}\Phi_{1,4}&\la_4^{n-1}\Phi_{2,4}&\la_4^{n}\Phi_{1,4}\\
\vdots&\vdots&\vdots&\vdots&\vdots&\vdots&\vdots&\vdots\\
\Phi_{1,2n-1}&\Phi_{2,2n-1}&\la_{2n-1}\Phi_{1,2n-1}&\la_{2n-1}\Phi_{2,2n-1}&\dots&\la_{2n-1}^{n-1}\Phi_{1,2n-1}&\la_{2n-1}^{n-1}\Phi_{2,2n-1}&\la_{2n-1}^{n}\Phi_{1,2n-1}\\
\Phi_{1,2n}&\Phi_{2,2n}&\la_{2n}\Phi_{1,2n}&\la_{2n}\Phi_{2,2n}&\dots&\la_{2n}^{n-1}\Phi_{1,2n}&\la_{2n}^{n-1}\Phi_{2,2n}&\la_{2n}^{n}\Phi_{1,2n}
\end{matrix}\right|\]}

{\tiny \[\notag(\T_n)_{21}&=&
\left|\begin{matrix}1&0&\la&0&\dots&\la^{n-1}&0&0\\
\Phi_{1,1}&\Phi_{2,1}&\la_1\Phi_{1,1}&\la_1\Phi_{2,1}&\dots&\la_1^{n-1}\Phi_{1,1}&\la_1^{n-1}\Phi_{2,1}&\la_1^{n}\Phi_{2,1}\\
\Phi_{1,2}&\Phi_{2,2}&\la_2\Phi_{1,2}&\la_2\Phi_{2,2}&\dots&\la_2^{n-1}\Phi_{1,2}&\la_2^{n-1}\Phi_{2,2}&\la_2^{n}\Phi_{2,2}\\
\Phi_{1,3}&\Phi_{2,3}&\la_3\Phi_{1,3}&\la_3\Phi_{2,3}&\dots&\la_3^{n-1}\Phi_{1,3}&\la_3^{n-1}\Phi_{2,3}&\la_3^{n}\Phi_{2,3}\\
\Phi_{1,4}&\Phi_{2,4}&\la_4\Phi_{1,4}&\la_4\Phi_{2,4}&\dots&\la_4^{n-1}\Phi_{1,4}&\la_4^{n-1}\Phi_{2,4}&\la_4^{n}\Phi_{2,4}\\
\vdots&\vdots&\vdots&\vdots&\vdots&\vdots&\vdots&\vdots\\
\Phi_{1,2n-1}&\Phi_{2,2n-1}&\la_{2n-1}\Phi_{1,2n-1}&\la_{2n-1}\Phi_{2,2n-1}&\dots&\la_{2n-1}^{n-1}\Phi_{1,2n-1}&\la_{2n-1}^{n-1}\Phi_{2,2n-1}&\la_{2n-1}^{n}\Phi_{2,2n-1}\\
\Phi_{1,2n}&\Phi_{2,2n}&\la_{2n}\Phi_{1,2n}&\la_{2n}\Phi_{2,2n}&\dots&\la_{2n}^{n-1}\Phi_{1,2n}&\la_{2n}^{n-1}\Phi_{2,2n}&\la_{2n}^{n}\Phi_{2,2n}
\end{matrix}\right|\]}

{\tiny \[\notag(\T_n)_{22}&=&
\left|\begin{matrix}0&1&0&\la&\dots&0&\la^{n-1}&\la^n\\
\Phi_{1,1}&\Phi_{2,1}&\la_1\Phi_{1,1}&\la_1\Phi_{2,1}&\dots&\la_1^{n-1}\Phi_{1,1}&\la_1^{n-1}\Phi_{2,1}&\la_1^{n}\Phi_{2,1}\\
\Phi_{1,2}&\Phi_{2,2}&\la_2\Phi_{1,2}&\la_2\Phi_{2,2}&\dots&\la_2^{n-1}\Phi_{1,2}&\la_2^{n-1}\Phi_{2,2}&\la_2^{n}\Phi_{2,2}\\
\Phi_{1,3}&\Phi_{2,3}&\la_3\Phi_{1,3}&\la_3\Phi_{2,3}&\dots&\la_3^{n-1}\Phi_{1,3}&\la_3^{n-1}\Phi_{2,3}&\la_3^{n}\Phi_{2,3}\\
\Phi_{1,4}&\Phi_{2,4}&\la_4\Phi_{1,4}&\la_4\Phi_{2,4}&\dots&\la_4^{n-1}\Phi_{1,4}&\la_4^{n-1}\Phi_{2,4}&\la_4^{n}\Phi_{2,4}\\
\vdots&\vdots&\vdots&\vdots&\vdots&\vdots&\vdots&\vdots\\
\Phi_{1,2n-1}&\Phi_{2,2n-1}&\la_{2n-1}\Phi_{1,2n-1}&\la_{2n-1}\Phi_{2,2n-1}&\dots&\la_{2n-1}^{n-1}\Phi_{1,2n-1}&\la_{2n-1}^{n-1}\Phi_{2,2n-1}&\la_{2n-1}^{n}\Phi_{2,2n-1}\\
\Phi_{1,2n}&\Phi_{2,2n}&\la_{2n}\Phi_{1,2n}&\la_{2n}\Phi_{2,2n}&\dots&\la_{2n}^{n-1}\Phi_{1,2n}&\la_{2n}^{n-1}\Phi_{2,2n}&\la_{2n}^{n}\Phi_{2,2n}
\end{matrix}\right|.\]}
\end{widetext}
\end{theorem}

The following identity can be found
\[T_n(\la;\la_1,\la_2,\la_3,\la_4,\dots,\la_{2n})|_{\la=\la_i}\left(\begin{matrix}\Phi_{1,i}\\
\Phi_{2,i}
\end{matrix}\right)=0,\]
where $i=1,2,3,\dots,2n.$ Similarly, the Darboux transformation for $(E, p,\eta)$ can  be constructed by using the following identities

\[
T_{nt}+T_nU&=&U^{[n]}T_n,\\
T_{nz}+T_nV&=&V^{[n]}T_n,
\]
which can be further simplified as
\[U_{0}^{[n]}=U_0+i[\sigma_3,t_{n-1}^{[n]}],\]
\[\label{V-1DT}V_{-1}^{[n]}=T_n|_{\la=-\om} V_{-1}T_n^{-1}|_{\la=-\om}.\]

The proof for Eq. \eqref{V-1DT} is quite complicated in general when compared to the proof of one and two fold Darboux transformation. However, this will become simple if we know the origin of $T_n$ \cite{Hedeterminant}. Because if we treat $n$-fold Darboux transformation as a generalization of $(n-1)$-Darboux transformation, the transformation will be the multiplications of $n$ one-fold Darboux transformations at $\la=-\om$. It is easy to prove that these multiplications can have a determinant representation as mentioned above.

The $n$-th new solution after the $n$-fold Darboux transformation of the H-MB system will be

\[E^{[n]}&=&E+2i(t_{n-1}^{[n]})_{12},\\ \notag
p^{[n]}&=&(2\eta(T_n)_{11}(T_n)_{12}
-p^*(T_n)_{12}(T_n)_{12}\\
&&+p(T_n)_{11}(T_n)_{11})\\  \notag
&&/((T_n)_{11}(T_n)_{22}-(T_n)_{12}(T_n)_{21})|_{\la=-\om},\\
\eta^{[n]}&=&(\eta((T_n)_{11}(T_n)_{22}+(T_n)_{12}(T_n)_{21})\\ \notag
&&-p^*(T_n)_{12}(T_n)_{22}
+p(T_n)_{11}(T_n)_{21})\\ \notag
&&/((T_n)_{11}(T_n)_{22}-(T_n)_{12}(T_n)_{21})|_{\la=-\om},
\]

where $\big(t_{n-1}^{[n]}\big)_{12}$ is the element at the first row and second column in the matrix $t_{n-1}^{[n]}$. So far, we discussed about the determinant construction of n-th Darboux transformation of the H-MB system. As an application of these transformations, soliton and positon solutions of the H-MB system will be constructed in the next section.

\section{Soliton solutions of the the H-MB system}\cleqn

In this section, having obtained the Daurboux transformation for our system, our next aim is to construct the one soliton solution of the H-MB system by assuming suitable seed solutions. We assume trivial seed solutions as $E=0, p=0, \eta=1$, then the linear system becomes

\[
\Phi_t&=&U\Phi,\\
\Phi_z&=&V\Phi,
\]
where

\[
\Phi&=&\left(\begin{matrix}\Phi_1\\ \Phi_2
\end{matrix}\right),\\
U&=&\left(\begin{matrix}-i\la& 0\\ 0& i\la
\end{matrix}\right),\\ \notag
V&=&\left(\begin{matrix}4i\beta\la^3-\alpha i\la^2 & 0\\ 0& -4\beta i\la^3+\alpha i\la^2
\end{matrix}\right)\\
&&+\frac{i}{\la+\om}\left(\begin{matrix}1 & 0\\ 0& -1
\end{matrix}\right).\]

From the above system, we construct the explicit eigenfunctions in the form

\[\notag
\Phi_1&=&e^{-i\la t+(4\beta i\la^3-\alpha i\la^2+\frac{i}{\la+\om})z+\frac{x_0+iy_0}2},\\ \notag
\Phi_2&=&e^{i\la t+(-4\beta i\la^3+\alpha i\la^2-\frac{i}{\la+\om})z-\frac{x_0+iy_0}2+i\theta},\]
where $x_0$, $y_0$ and $\theta$ are all arbitrarily fixed real constants. Substituting these two eigenfunctions into the one-fold Darboux transformation Eq. \eqref{Edetail}, Eq. \eqref{pdetail}, Eq. \eqref{etadetail} and choosing $\la=\alpha_1+I \beta_1$, $x_0=0,\,y_0=0, \theta=0$, then the following  form of soliton solutions are obtained:

\[E&=&2 \beta_1 e^{-2 i \frac{C}{B}}sech(2 \beta_1 \frac{A}{B}),\\ \notag
p&=&\frac{ i \beta_1 ((\alpha_1 -i \beta_1+\omega) e^{-2 \frac{D}{B}}+(\alpha_1 +\beta_1 i+\omega )e^{-2 \frac{F}{B}})}{ B}\\
&&sech^2(2 \beta_1 \frac{A}{B}),\\
\eta
&=&1-\frac{2\beta_1^2}{B}sech^2(2 \beta_1 \frac{A}{B}),
\]
where $A,B,C,D,F$ are explicitly given in Appendix I.

If we choose $\alpha=1,\, \beta=0$, the one soliton solution is just the soliton solution of Eq. (15) of NLS-MB system mentioned in \cite{heTheor} with $\alpha_1=-\rho_1,\,\beta_1=\nu_1.$ Similarly, substituting these two eigenfunctions into the one-fold Darboux transformation eq.\eqref{Edetail}, eq.\eqref{pdetail}, eq.\eqref{etadetail} and  choosing $\alpha=2,\,\beta=-1$, then the one-solition solutions of the classical H-MB system can be obtained whose evolution is shown in Fig.\ref{1solitonH-MB}, which clearly indicates that $E$ and $p$ are bright solitons because their waves are under the flat non-vanishing plane whereas $\eta$ is a dark soliton.

Now let us discuss the construction of the two-soliton solutions of the H-MB system. For this purpose, we have to use two spectral parameters $\lambda_1=\alpha_1+i\beta_1$ and $\lambda_2=\alpha_2+ i\beta_2$. After the second Darboux transformation, we can construct the two solition solutions.  As the general form of two soliton solution is quite tedious in nature, for simplicity, we are giving only the two soliton solution of E with $\omega=1.5,\alpha_1=0.5,\beta_1=1,\alpha_2=1,\beta_2=1.5,\alpha=2$ and $\beta=-1$,

\begin{eqnarray*}
&&E_{2-sol}=\\&&-i(17e^{-0.01176470588i(85t-1258z+1815iz+255it)}\\
&&-(36+18i)e^{-0.01176470588i(170t-4385z+204iz+170it)}
\\
&&-7e^{0.01176470588i(-85t+1258z+1815iz+255it)}\\
&&
+(18i-12)e^{0.01176470588i(-170t+4385z+204iz+170it)}\\
&&-16ie^{0.01176470588i(-85t+1258z+1815iz+255it)}\\
&&
+20ie^{-0.01176470588i(85t-1258z+1815iz+255it)})\\
&&
/(-24cos(0.9999999998t-36.78823529z)\\
&&+26cosh(t+18.95294118z)\\
&&+2cosh(-5t-23.75294118z).
\end{eqnarray*}

We also constructed the two soliton solution for p and $\eta$ in a similar manner. For completeness, instead of giving complicated forms of p and $\eta$, the graphical representation of them is shown in Fig.\ref{2solitonH-MB}.

\section{Bright and dark positon solutions of the H-MB system}\cleqn

In the case of two soliton solution constructed above, if the
second spectral parameter $\la_2$ is assumed to be close to the
first spectral parameter $\la_1$, and doing the Taylor expansion of
wave function to first order up to $\la_1$ will lead to a new kind
of solution which is called as degenerate soliton\cite{hehrwnls} -
smooth positon solution. ``Positon" was coined by
Matvee \cite{Matveev92pla,Matveev92pla2, Matveev02TMP} for the Korteweg-de Vries (KdV)
equation by the same limiting approach. Note that the positon of the
KdV is a singular solution. For the construction of positon solutions,
 following four linear functions are needed to construct the second
Darboux transformation and to generate the positon solutions,

\begin{eqnarray*}
\Phi_{1,1}&=&e^{-i\la_1 t+(4\beta i\la_1^3-\alpha i\la_1^2+\frac{i}{\la_1+\om})z+\frac{x_0+iy_0}2}  ,\\
\Phi_{2,1}&=&e^{i\la_1 t+(-4\beta i\la_1^3+\alpha i\la_1^2-\frac{i}{\la_1+\om})z-\frac{x_0+iy_0}2+i\theta} ,\\
\Phi_{1,3}&=&e^{-i\la_3 t+(4\beta i\la_3^3-\alpha i\la_3^2+\frac{i}{\la_3+\om})z+\frac{x_0+iy_0}2},\\
\Phi_{2,3}&=&e^{i\la_3 t+(-4\beta i\la_3^3+\alpha i\la_3^2-\frac{i}{\la_3+\om})z-\frac{x_0+iy_0}2+i\theta}.
\end{eqnarray*}

Now we take $\la_3=\la_1+\ep(1+i)$ and use the  Taylor expansion of wave function $\phi_3$ and $\phi_4$ up to first order of $\ep$ in terms of $\la_1$.  For example, choosing $\omega=1.5,\,\alpha_1=0.5,\,\beta_1=1,\alpha=2\,\,and\,\, \beta=-1$, the positon solutions $E_{p}$  is constructed in the form

\begin{eqnarray*}&&E_{p}=8ie^{14.8iz-it}\times\\
&&(1582zcosh(2t+2.40z)
-688izsinh(2t+2.4z)\\
&&-50icosh(2t+2.4z)+100itsinh(2t+2.4z))/\\
&&(400t^2+119296z^2-5504tz+50+50cosh(4.8z+4t)).
\end{eqnarray*}

In this case, the pictorial representation of the positon solutions $(E_{p},p_{p},\eta_{p})$ of the H-MB system is shown in Fig.\ref{Hpositon}.

From the above figures, we observe that two peaks of the positon solutions are at same height which is different from  the two-soliton solution. Meanwhile the two waves depart at a relatively less speed after their collision. This is also different from the two-solitons, which depart at a fixed speed.
Here again, we find that $E$ and $p$ are bright positon solutions whereas $\eta$ is a dark positon in all three cases discussed above.

\section{Bright and dark breather solutions of the H-MB system}
In the last two sections, soliton solutions and positon solutions have been generated for the H-MB system. In this section, we will now focus on a new and different kind of solution which is also derived from periodic solutions through Darboux transformation. The resulting periodic solutions can be called breather solutions. Now, let us assume the seed solutions as $E=d e^{i\rho}, p=if E, \eta=1,\rho=az+bt$, which admits the constraint in the form
\[2a+\alpha b^2-2 \alpha d^2+2 \beta b^3-12\beta d^2 b-4 f=0, \\
 fb-2\omega f+2=0.\]

Defining
 \[R:=\sqrt{-b^2-4b\lambda-4\lambda^2-4d^2},\]  the following wave function $\Phi=\left(\begin{matrix}\phi_1\\
\phi_2\end{matrix}\right)$ is obtained in terms of $R$ as

\[\phi_1(R) &:=& e^{\frac{i( P_1 z
+ Q_1 t)}{4(b-2 \omega)(\lambda+\omega)}},\\
\phi_2(R)& :=& \frac{1}{2d} (b i+2 i \lambda+R)e^{\frac {i(P_2 z
+ Q_2t)}{4(b-2 \omega)(\lambda+\omega)}}.\]
where
$P_1,Q_1,P_2,Q_2$ are polynomials independent of $t$ and $z$ and their complete expressions are given in Appendix II.

If we use the above two wave functions to construct the two new functions $h_1$ and $h_2$ as in \cite{heTheor}, then the resulting new solutions obtained through Darboux transformation are found to have no meaning. Therefore, we would like to construct more complicated but physically meaningful solutions in the following part. By combining these two wave functions, we derive the new functions $h_1$ and $h_2$ as follows
\[h_1:=g_1(\la)-g_1(\la^*)^*, h_2:=g_2(\la)+g_1(\la^*)^*,\]
where
\[g_1:=\phi_1(R)+\phi_1(-R), g_2:=\phi_2(R)+\phi_2(-R).\]

It can be proved that $h_1$ and $h_2$ are also the solutions of the Lax equation with $\la:=\alpha_1+i\beta_1$. Using these two wave functions $h_1$ and $h_2$ in the one-fold Darboux transformation will lead to the construction of breather solutions of the H-MB system. To simplify the calculations, we take $b=-2\alpha_1$ and use the second Darboux transformation discussed in the last section, then the final form of the breather solution $E_{b}$ is obtained in the form

\begin{eqnarray}\notag
E_{b}&=&( d^2(e^{i \frac{A_b(w)}{Y}}+e^{\frac{A_b(-w)}{Y} i})+2 d\beta_1( e^{i\frac{C_b^*}{Y}}+ e^{\frac{C_b}{Y} i}) \\ \notag
&& -d \beta_1( e^{ \frac{D_b}{X}i}+ e^{ \frac{D_b^*}{X}i})+2 \beta_1(w-\beta_1)e^{\frac{F_b(w)}{Y} i}\\ \label{Ebrea}
&& -2 \beta_1(w+\beta_1) e^{\frac{F_b(-w)}{Y} i})/\\ \notag
&&(
2d cosh (2 i w \beta_1 z \frac{G_b}{Y})-2 \beta_1 cosh(2 w \frac{H_b}{Y})),\end{eqnarray}

where

\[w=\sqrt{\beta_1^2-d^2},\]
\[X=(\alpha_1+\beta_1 i+\omega)(-\alpha_1+\beta_1 i-\omega),\]
\[Y=(\alpha_1+\omega)(\alpha_1+\beta_1 i+\omega)(-\alpha_1+\beta_1 i-\omega),\]
and
$A_b(w),C_b,D_b,F_b(w),$ and $G_b,H_b$ are polynomials of $t, z,\omega,w,a,b,d,\alpha,\beta,\alpha_1,\beta_1$ which are defined in Appendix-II.

Similarly, the breather form of $p$ and $\eta$ can be constructed. For  example, after taking values \[\notag \omega=0.5,d=0.5,\alpha_1=-1,\beta_1=1,\alpha=2,\beta=-1,\\\] the breather solution  of the H-MB system is plotted in Fig.\ref{Hbreather} and Fig.\ref{Hbreatherdensity}.

Similarly, for the next choice of parameters $\omega=1.5,\,\alpha_1=0.5,\,\beta_1=1,\,\alpha=0,\, \beta=1$, the breather solutions of the complex modified
Korteweg-de Vries (CMKdV)-MB system are obtained. The picture of breather solutions is shown in Fig.\ref{01breather}. In addition to the above observation, we also find how the values of $(\alpha, \beta)$ changes the direction of breather solution $\eta$ in the $(t, z)$ plane, see Fig.\ref{breatherdirection}. To the best of our knowledge, we observe this effect for the first time.

Having constructed bright breathers for $E$ and $p$ and a dark breather for $\eta$, in the next section, our aim is to discuss the construction of rogue wave solutions of the H-MB system which is in fact one single period of breather solutions.

\section{Bright and dark rogue waves in the H-MB system}\cleqn

In this section, using the limit method of the NLS equation, we construct the rogue wave solutions of the H-MB system \cite{hehrwnls}. This kind of solution only appears in some special regions of time and distance and then will be drowned in one fixed non-vanishing plane. If we do the Taylor expansion to the breather solution \eqref{Ebrea} around $\beta_1=d$, one rogue wave solution of $E$ will be obtained and rogue waves for $p$
 and $\eta$ can also be constructed in a similar way. In the following,
 for brevity, we only report the rogue wave for $E_{r}$ in the form

\[\notag E_{r}&=& \frac{\bar A(t,z) d}{\bar B(t,z)}\exp( \frac{i}{\alpha_1+\omega}(8 z \beta \alpha_1^4+8 z \beta \alpha_1^3 \omega\\ \notag
&&-2 z \alpha \alpha_1^2 \omega+2 z
-2 \alpha_1 t \omega+z \alpha d^2 \omega-12 z \beta d^2 \alpha_1 \omega\\
&&-2 z \alpha \alpha_1^3-2 \alpha_1^2 t+z \alpha d^2 \alpha_1-12 z \beta d^2 \alpha_1^2
)) ,\]

where
$\bar A(t,z),\bar B(t,z)$ are polynomials of $t,z,\omega,w,a,b,d,\alpha,\beta,\alpha_1,\beta_1$ which are defined in Appendix-III.

When we take $\omega=0.5,\,d=0.5,\,\alpha_1=-1,\,\beta_1=1,\,\alpha=2,\,\beta=-1$, the final form of the rogue wave  solutions will be

    \[\label{1roguevelocity}E_{r}&=&(\frac{8(1-3i z)}{4+(2t+17z)^2+36z^2}-\frac12)e^{0.5i(-5z+4t)},\\
   \label{1roguep}\notag p_{r}&=& e^{0.5i(-5z+4t)}(1031493073iz^4+10312500iz^2\\ \notag
 &&-5312500z-625000t-152343760z^3\\
 \notag &&-1875000t^2z-31875000tz^2\\ \notag
 &&
 +70546875it^2z^2+5312500it^3z+431640580itz^3\\ \notag
 &&+1875000itz+156250it^4-156250i)/(156250t^4\\ \notag
 && +25390625z^2+312500t^2
 +431640580tz^3\\
 \notag &&+70546900t^2z^2+5312500t^3z\\
 &&+156250+5312500tz+1031493636z^4),\\
 \notag\eta_{r}&=& (16t^4+64t^2+7224t^2z^2+105625z^4+2992z^2\\ \notag
 &&+44200tz^3-16+544t^3z+896zt)\\ \notag
 &&/(16t^4+32t^2+7224t^2z^2+105625z^4\\ \notag
 &&+2600z^2+44200tz^3+16+544t^3z+544zt).\\
 \label{1rogueeta}\]

From Eq. \eqref{1roguevelocity}, it is clearly observed that the height of the background of $|E_r|^2$ is ¼ and the orders of the numerators and denominators of $ p_{r}$ and $ \eta_{r}$ are four. Because of these reasons, the graphs of the rogue waves for $ p_{r}$ and $ \eta_{r}$ have double peaks which are shown in Fig.\ref{H-MB-rogue} and the corresponding density graph is plotted in Fig.\ref{H-MB-rogue-density}.

For further understanding of our observations, we enlarge the above density in Fig.\ref{H-MB-rogue-density}, some zoomed portions of the above figures are clearly shown in Fig.\ref{H-MB-rogue-density-close}. From the graph of $|p|^2$ shown in Fig.\ref{H-MB-rogue-density-close}, we find one cave appears on top of the single peak  with two caves on both sides of the peak.

 To realize the significance of the different parameters $\alpha$ and $\beta$,  we also consider the case, when $\alpha=0,\, \beta=1$, i.e., the CMKdV-MB system, see Fig.\ref{01-MB-rogue}. From the figures, we also find that the parameters $\alpha$ and $\beta$  will change the shape, pulse width, etc., of the  rogue wave. Therefore, in the following, we fix time $t$ and distance $z$ to see the role of  parameters  $\alpha$ and $\beta$ and their impact on rogue wave dynamics.

We keep $\alpha$ and $\beta$ as arbitrary parameters and choose  $\omega=0.5,\,d=0.5,\,\alpha_1=-1,\,\beta_1=1,\,t=1,\,z=1.$
This is also a rogue wave whose graph is portrayed in Fig.\ref{t=1z=1alphabeta-rogue}.  Here we provide only the specific form of the rogue wave solution $E_{r\alpha\beta}$ as

\begin{eqnarray}\notag E_{r\alpha\beta}&=& e^{-\frac{i}4(8+20\beta+7\alpha)}(40\alpha-17\alpha^2+156\beta-40\\ \notag
&&-585\beta^2-192\alpha\beta+i(32+8\alpha+96\beta))/\\ \notag
&&(1170\beta^2-80\alpha+34\alpha^2-312\beta+112+384\alpha\beta).\\\label{rogalphabeta}
\end{eqnarray}

This implies that after fixing the values of $t$ and $z$, the solution depending on parameters $\alpha,\,\beta$ is also in the form of a rogue wave, which is observed for the first time. This will give us some idea about how to modify the parameters $\alpha$ and $\beta$ to visualize our theoretical results in terms of the experimental results  in optics. From the above, one can easily conclude that $E$ and $p$ are bright rogue waves and $\eta$ is a dark rogue wave.

All the  solutions mentioned above including positons and rogue wave solutions are indeed solutions of the H-MB system  which are verified by using the symbolic computation software MAPLE.

\section{Conclusion and Discussions}

In this paper, after a  suitable choice of self-steepening and self-frequency shift effects, we have derived the Darboux transformation of the H-MB system which is governed by ultra-short pulse propagation through an erbium doped nonlinear optical waveguide and further generalized it to the matrix form of an $n$-fold Darboux transformation, which implies the determinant representation of $(E^{[n]},p^{[n]}, \eta^{[n]})$ generated from the known solution $(E,p, \eta )$. By choosing some special eigenvalues $\la_{2n-1}=\la_{2n}^*$ and eigenfunctions using the reduction conditions $\Phi_{1,2n-1}=\Phi_{2,2n}^*,\ \ \Phi_{2,2n-1}=-\Phi_{1,2n}^*$, the determinant representation of $(E^{[n]},p^{[n]}, \eta^{[n]})$ provided some new solutions of the H-MB system. As examples, soliton solutions, breather solutions, and rogue wave solutions of the H-MB system have been constructed explicitly by using  the Darboux transformation from trivial and periodic seed solutions.  The rogue waves show interesting characteristics which might attract physicists to observe them in experiments with higher order optical effects in the femtosecond regime. The interesting characteristics obtained contain the following two sides: (i) The rogue wave solution for $p,\,\eta$ is surprisingly found by us to have two peaks because the order of the numerator and denominator of $p,\,\eta$ in eq.\eqref{1roguep} and eq.\eqref{1rogueeta} is four and (ii) after fixing the time and spatial parameter and by changing other two unknown parameters $\alpha$ and $\beta$, we find a rogue wave shape also arises out as shown in eq.\eqref{rogalphabeta}. This is the first time that this phenomenon is obtained to the best of our knowledge. Still, there are a few interesting questions which are still unclear. For example, the physical interpretations and observation of higher-order positon solutions, the role of higher-order rogue waves solutions and their applications in physics, in particular, the connection between rogue wave solutions and supercontinuum generation through modulation instability or soliton fission, etc., with higher-order optical effects.

{\bf Acknowledgments} {\noindent \small  This work is supported by
the National Natural Science Foundation of China under Grant No.11201251,
 the Natural Science Foundation of Zhejiang
Province under Grant No. LY12A01007. J.S.He is supported by  the
National Natural Science Foundation of China under Grant No.
10971109 and NO. 11271210, and the K.C. Wong Magna Fund at Ningbo University. K.
Porseizan wishes to thank the DST, DAE-BRNS, UGC, and CSIR,
Government of India, for financial support through major
projects. }
\begin{widetext}

\section{\textbf{Appendices}}
\begin{center}
\appendix{\textbf{Appendix I}}:
In this appendix, we are providing  explicit expression for A, B, C, D, and F
\end{center}

\begin{eqnarray*}A&=&-24 z \beta \alpha_1^3 \omega+4 z \alpha \alpha_1^2 \omega-8 z \beta \alpha_1^2 \beta_1^2+2 z \alpha \alpha_1 \beta_1^2+4 \beta_1^2 z \beta \omega^2+2 z \alpha \alpha_1 \omega^2\\
&&-12 z \beta \alpha_1^2 \omega^2+z+8 z \beta \alpha_1 \beta_1^2 \omega+t \alpha_1^2+t \beta_1^2+t \omega^2+2 t \alpha_1 \omega-12 z \beta \alpha_1^4+2 z \alpha \alpha_1^3+4 z \beta \beta_1^4,\\
B&=&\alpha_1^2+2\alpha_1\omega+\omega^2+\beta_1^2,\\
C&=&2 t \alpha_1^2 \omega+t \alpha_1 \omega^2+t \beta_1^2 \alpha_1-4 z \beta \alpha_1^5+z \alpha \alpha_1^4-z \alpha \beta_1^4-z \alpha_1+t \alpha_1^3\\
&&+24 z \beta \alpha_1^2 \beta_1^2 \omega+12 z \beta \alpha_1 \beta_1^2 \omega^2-2 z \alpha \alpha_1 \beta_1^2 \omega-8 z \beta \alpha_1^4 \omega-4 z \beta \alpha_1^3 \omega^2\\
&&+8 z \beta \alpha_1^3 \beta_1^2+12 z \beta \beta_1^4 \alpha_1+2 z \alpha \alpha_1^3 \omega+z \alpha \alpha_1^2 \omega^2-z \alpha \beta_1^2 \omega^2-z \omega,\end{eqnarray*}

\begin{eqnarray*}
D&:=&-24 \beta_1 z \beta \alpha_1^3 \omega+4 \beta_1 z \alpha \alpha_1^2 \omega+2 \beta_1 z \alpha \alpha_1 \omega^2-12 \beta_1 z \beta \alpha_1^2 \omega^2+8 \beta_1^3 z \beta \alpha_1 \omega+\beta_1 z+\beta_1^3 t+4 \beta_1^5 z \beta\\
&&+\beta_1 t \omega^2+\beta_1 t \alpha_1^2-8 \beta_1^3 z \beta \alpha_1^2+2 \beta_1^3 z \alpha \alpha_1+4 \beta_1^3 z \beta \omega^2+2 \beta_1 t \alpha_1 \omega-12 \beta_1 z \beta \alpha_1^4+2 \beta_1 z \alpha \alpha_1^3\\&&+i(t \beta_1^2 \alpha_1 +z \alpha \alpha_1^4 - z \alpha_1- z \omega+t \alpha_1^3 +t \alpha_1 \omega^2 +2  t \alpha_1^2 \omega- z \alpha \beta_1^2 \omega^2\\
&&-4  z \beta \alpha_1^3 \omega^2-8  z \beta \alpha_1^4 \omega-4  z \beta \alpha_1^5- z \alpha \beta_1^4-2  z \alpha \alpha_1 \beta_1^2 \omega\\
&&+8  z \beta \alpha_1^3 \beta_1^2+12  z \beta \beta_1^4 \alpha_1+2  z \alpha \alpha_1^3 \omega+24  z \beta \alpha_1^2 \beta_1^2 \omega+12  z \beta \alpha_1 \beta_1^2 \omega^2+z \alpha \alpha_1^2 \omega^2 ),
\end{eqnarray*}

\begin{eqnarray*}F&:=&24 \beta_1 z \beta \alpha_1^3 \omega-4 \beta_1 z \alpha \alpha_1^2 \omega-2 \beta_1 z \alpha \alpha_1 \omega^2+12 \beta_1 z \beta \alpha_1^2 \omega^2-8 \beta_1^3 z \beta \alpha_1 \omega-\beta_1 z-\beta_1^3 t-4 \beta_1^5 z \beta\\
&&-\beta_1 t \omega^2-\beta_1 t \alpha_1^2+8 \beta_1^3 z \beta \alpha_1^2-2 \beta_1^3 z \alpha \alpha_1-4 \beta_1^3 z \beta \omega^2-2 \beta_1 t \alpha_1 \omega+12 \beta_1 z \beta \alpha_1^4-2 \beta_1 z \alpha \alpha_1^3\\
&&+i(t \beta_1^2 \alpha_1 +z \alpha \alpha_1^4 - z \alpha_1- z \omega+t \alpha_1^3 +t \alpha_1 \omega^2 +2  t \alpha_1^2 \omega- z \alpha \beta_1^2 \omega^2\\
&&-4  z \beta \alpha_1^3 \omega^2-8  z \beta \alpha_1^4 \omega-4  z \beta \alpha_1^5- z \alpha \beta_1^4-2  z \alpha \alpha_1 \beta_1^2 \omega\\
&&+8  z \beta \alpha_1^3 \beta_1^2+12  z \beta \beta_1^4 \alpha_1+2  z \alpha \alpha_1^3 \omega+24  z \beta \alpha_1^2 \beta_1^2 \omega+12  z \beta \alpha_1 \beta_1^2 \omega^2+z \alpha \alpha_1^2 \omega^2),
\end{eqnarray*}

\begin{center}
\appendix{\textbf{Appendix II}}:
In this appendix, we are furnishing the expression for $P_1,Q_1,P_2,Q_2$
\end{center}

\begin{eqnarray*}P_1&=&12 \beta d^2 b^2 \lambda-4 \alpha d^2 \lambda \omega+4 \beta b^3 \lambda \omega+2 \alpha d^2 b \omega+2 \alpha d^2 b \lambda\\&&
+2 \alpha b^2 \lambda \omega-24 \beta d^2 b \omega^2+12 \beta d^2 b^2 \omega-8 \lambda-8 \omega-4 \alpha d^2 \omega^2-2 \beta b^4 \lambda\\&&
-2 \beta b^4 \omega+2 \alpha b^2 \omega^2+4 \beta b^3 \omega^2-\alpha b^3 \lambda-\alpha b^3 \omega-24 \beta d^2 b \lambda \omega+2 i \omega \beta R b^3\\&&
+2 i \lambda \beta R b^3+\lambda \alpha R b^2 i+\omega \alpha R b^2 i+16 i \lambda^2 \omega \beta R b+4 i R+8 i \beta \lambda b \omega^2 R\\&&
-8 i \beta \lambda b^2 \omega R-4 i d^2 \beta \omega R b-4 i d^2 \beta \lambda R b-4 i \alpha b \lambda R \omega+8 i d^2 \beta \lambda R \omega\\&&
-4 i \beta b^2 \omega^2 R-16 i \beta \lambda^3 R \omega-2 i \alpha \lambda^2 R b-2 i \alpha b \omega^2 R-16 i \beta \lambda^2 \omega^2 R\\&&
-4 i \beta \lambda^2 b^2 R+4 i \alpha \lambda^2 R \omega+8 i \beta \lambda^3 R b+8 i d^2 \beta \omega^2 R+4 i \alpha \lambda \omega^2 R,\\
Q_1&=& 2 b^2\omega-4 b  \lambda \omega-4 b  \omega^2+2 b^2 \lambda-2 i  R b \omega+4 i  R \omega \lambda+4 i  R \omega^2-2 i  R b \lambda,
\end{eqnarray*}
\begin{eqnarray*}
P_2&=&-12 \beta d^2 b^2 \lambda+4 \alpha d^2 \lambda \omega-4 \beta b^3 \lambda \omega-2 \alpha d^2 b \omega
-2 \alpha d^2 b \lambda-2 \alpha b^2 \lambda \omega+24 \beta d^2 b \omega^2\\&&
-12 \beta d^2 b^2 \omega+8 \lambda+8 \omega+4 \alpha d^2 \omega^2
+2 \beta b^4 \lambda+2 \beta b^4 \omega
-2 \alpha b^2 \omega^2-4 \beta b^3 \omega^2\\&&
+\alpha b^3 \lambda+\alpha b^3 \omega+24 \beta d^2 b \lambda \omega
+2 i \omega \beta R b^3+2 i \lambda \beta R b^3+\lambda \alpha R b^2 i
+\omega \alpha R b^2 i\\&&
+16 i \lambda^2 \omega \beta R b+4 i R
+8 i \beta \lambda b \omega^2 R-8 i \beta \lambda b^2 \omega R-4 i d^2 \beta \omega R b-4 i d^2 \beta \lambda R b\\&&
-4 i \alpha b \lambda R \omega
+8 i d^2 \beta \lambda R \omega-4 i \beta b^2 \omega^2 R-16 i \beta \lambda^3 R \omega-2 i \alpha \lambda^2 R b-2 i \alpha b \omega^2 R\\&&
-16 i \beta \lambda^2 \omega^2 R-4 i \beta \lambda^2 b^2 R+4 i \alpha \lambda^2 R \omega+8 i \beta \lambda^3 R b+8 i d^2 \beta \omega^2 R
+4 i \alpha \lambda \omega^2 R,\\
Q_2&=&4 b \lambda \omega+4 b \omega^2-2 b^2  \omega-2 b^2  \lambda-2 i  R b \omega+4 i  R \omega \lambda
+4 i  R \omega^2-2 i  R b \lambda,
\end{eqnarray*}

\begin{center}
\appendix{\textbf{Appendix III}}:
In this appendix, we are providing the expression for \\$A_b(w),\,C_b,\,D_b,\,F_b(w),\,G_b\,\,H_b, \bar A(t,z)\, and \, \bar B(t,z)$
\end{center}

\begin{eqnarray*}A_b&=&-2 w \beta_1 z
-72 z \beta \alpha_1^3 \omega w \beta_1-72 z \beta \alpha_1^2 \omega^2 w \beta_1+6 z \alpha \alpha_1^2 \omega w  \beta_1
+6 z \alpha \alpha_1 \omega^2 w \beta_1\\&&
-24 z \beta \omega w \alpha_1 \beta_1^3-24 z \beta \alpha_1 \omega^3 w \beta_1
+2 z \alpha \alpha_1^3 w \beta_1-24 z \beta \alpha_1^4 w \beta_1+2 z \alpha \omega^3 w \beta_1
-24 z \beta w \alpha_1^2 \beta_1^3\\&&
+2 z \alpha w \alpha_1 \beta_1^3+2 z \alpha w \omega \beta_1^3-(a z \alpha_1^3+a z \omega^3+b t \alpha_1^3+b t \omega^3+3 a z \alpha_1^2 \omega\\ &&
+3 a z \alpha_1 \omega^2+3 b t \alpha_1^2 \omega+3 b t \alpha_1 \omega^2+a z \alpha_1 \beta_1^2+a z \omega \beta_1^2+b t \alpha_1 \beta_1^2+b t \omega \beta_1^2),
\end{eqnarray*}
\begin{eqnarray*}
C_b&=&-2 z \omega^2-2 z \beta_1^2+2 \alpha_1^4 t-2 z \alpha_1^2-4 z \omega  \alpha_1+6 \alpha_1^3 t \omega+2 z \alpha \alpha_1^5-8 z \beta \alpha_1^6\\
&&+6 \alpha_1^2 t \omega^2+2 \alpha_1 t \omega^3-z \alpha d^2 \omega^3-z \alpha d^2 \alpha_1^3+6 z \alpha \alpha_1^4 \omega+6 z \alpha \alpha_1^3 \omega^2\\&&
+2 z \alpha \alpha_1^2 \omega^3-24 z \beta \alpha_1^5 \omega-24 z \beta \alpha_1^4 \omega^2-8 z  \beta \alpha_1^3 \omega^3+12 z \beta d^2 \alpha_1^4\\&&
-3 z \alpha d^2 \omega^2 \alpha_1-3 z \alpha d^2 \alpha_1^2 \omega+36 z \beta d^2 \alpha_1^3 \omega+36 z \beta d^2  \alpha_1^2 \omega^2\\&&
+12 z \beta d^2 \alpha_1 \omega^3-8 z \beta \alpha_1^4 \beta_1^2+2 z \alpha \alpha_1^3 \beta_1^2+2 \alpha_1 t \omega \beta_1^2+2 \alpha_1^2 t \beta_1^2\\
&&+4 i z d^2 \beta w \omega \beta_1^2+4 i z \alpha \alpha_1 w \omega \beta_1^2+4 i z d^2 \beta w \alpha_1  \beta_1^2+24 i z \beta \alpha_1 \omega^2 w \beta_1^2\\&&
+12 i z d^2 \beta \omega^2 w \alpha_1+12 i z d^2 \beta w \alpha_1^2 \omega+2 i t w \omega \beta_1^2+2 i t w \alpha_1 \beta_1^2\\&&
+6 i t w \omega^2 \alpha_1+6 i t w \alpha_1^2 \omega+4 i z \alpha \alpha_1^4 w-24 i z \beta \alpha_1^5 w+12 z \beta d^2 \alpha_1^2 \\&&
\beta_1^2+2 z \alpha \alpha_1^2 \omega \beta_1^2-z \alpha d^2 \alpha_1 \beta_1^2-z \alpha d^2 \omega \beta_1^2-8 z \beta \alpha_1^3 \omega \beta_1^2\\&&
+12 z \beta d^2  \alpha_1 \omega \beta_1^2-72 i z \beta \alpha_1^4 \omega w-72 i z \beta \alpha_1^3 \omega^2 w-24 i z \beta \alpha_1^2 \omega^3 w\\&&
+4 i z \alpha w \alpha_1^2 \beta_1^2+12 i z \alpha \alpha_1^3 \omega w+12 i z \alpha \alpha_1^2 \omega^2 w+4 i z \alpha \alpha_1 \omega^3 w\\&&
+4 i z d^2 \beta \omega^3 w+4 i z d^2 \beta w \alpha_1^3-16 i z \beta \alpha_1^3 w \beta_1^2+8 i z \beta \omega^3 w \beta_1^2+8 i z \beta w \alpha_1 \\&& \beta_1^4+8 i z \beta w \omega \beta_1^4+2 i z w \alpha_1+2 i z w \omega+2 i t w \omega^3+2 i t w \alpha_1^3,
\end{eqnarray*}
\begin{eqnarray*}
D_b&=&8 i z \alpha \alpha_1^2 \omega w+4 i z \alpha \alpha_1 \omega^2 w+4 i z d^2 \beta \omega^2 w+4 i z d^2 \beta w \alpha_1^2\\&&
+8 i z \beta \omega^2 w \beta_1^2+2 i z w+4 i z \alpha \alpha_1^3 w+4 i t w \alpha_1 \omega-48 i z \beta \alpha_1^3 \omega w-24 i z \beta \alpha_1^2 \omega^2 w\\&&
+2 i t w \omega^2+2 i t  w \alpha_1^2-24 i z \beta \alpha_1^4 w+8 i z d^2 \beta \omega w \alpha_1-a z \alpha_1^2-a z \omega^2-b t \alpha_1^2\\&&
-b t \omega^2-2 a z \alpha_1 \omega-2 b t \alpha_1 \omega-a z \beta_1^2-b t \beta_1^2+8 i z \beta w \beta_1^4+16 i z \beta \omega w \alpha_1 \beta_1^2\\&&
+4 i z \alpha w \alpha_1 \beta_1^2-16 i z \beta \alpha_1^2 w \beta_1^2+4 i z d^2 \beta  w \beta_1^2+2 i t w \beta_1^2,\\
F_b&=&-2 z \omega^2-2 z \beta_1^2-2 w \beta_1 z+2 \alpha_1^4 t-2 z \alpha_1^2-4 z \omega \alpha_1+6 \alpha_1^3 t \omega+2 z \alpha \\&& \alpha_1^5-8 z \beta \alpha_1^6+6 \alpha_1^2 t \omega^2+2 \alpha_1 t \omega^3-z \alpha d^2 \omega^3-z \alpha d^2 \alpha_1^3+6 z \alpha \alpha_1^4 \omega+6 z \alpha \\&&
 \alpha_1^3 \omega^2+2 z \alpha \alpha_1^2 \omega^3-24 z \beta \alpha_1^5 \omega-24 z \beta \alpha_1^4 \omega^2-8 z \beta \alpha_1^3 \omega^3+12 z \beta d^2 \\&& \alpha_1^4-3 z \alpha d^2 \omega^2 \alpha_1-3 z \alpha d^2 \alpha_1^2 \omega+36 z \beta d^2 \alpha_1^3 \omega+36 z \beta d^2 \alpha_1^2 \omega^2+12 z \beta d^2 \alpha_1 \\&&
 \omega^3-72 z \beta \alpha_1^3 \omega w \beta_1-72 z \beta \alpha_1^2 \omega^2 w \beta_1+6 z \alpha \alpha_1^2 \omega
 w \beta_1+6 z \alpha \alpha_1 \omega^2 w \beta_1\\&&
 -24 z \beta \omega w \alpha_1 \beta_1^3-24 z \beta \alpha_1  \omega^3 w \beta_1+2 z \alpha \alpha_1^3 w \beta_1-24 z \beta \alpha_1^4 w \beta_1+2 z \alpha \omega^3 \\&& w \beta_1-24 z \beta w \alpha_1^2 \beta_1^3+2 z \alpha w \alpha_1 \beta_1^3+2 z \alpha w \\&& \omega \beta_1^3-8 z \beta \alpha_1^4 \beta_1^2+2 z \alpha \alpha_1^3 \beta_1^2+2 \alpha_1 t \omega \beta_1^2+2 \alpha_1^2 t \beta_1^2+12 z \beta d^2 \alpha_1^2 \beta_1^2+2 z\\&& \alpha \alpha_1^2 \omega \beta_1^2-z \alpha d^2 \alpha_1 \beta_1^2-z \alpha d^2 \omega \beta_1^2-8 z \beta \alpha_1^3 \omega \beta_1^2+12 z \beta d^2 \alpha_1 \omega \beta_1^2,
 \end{eqnarray*}
\begin{eqnarray*}
 G_b&=&-1-12 \beta \alpha_1^4+\alpha \alpha_1^3+3 \alpha  \alpha_1^2 \omega-36 \beta \alpha_1^3 \omega+\alpha \omega^3+\alpha \omega \beta_1^2
-36 \alpha_1^2 \beta \omega^2\\&&-12 \alpha_1^2 \beta \beta_1^2+\alpha_1 \alpha \beta_1^2+3  \alpha_1 \alpha \omega^2-12 \alpha_1 \omega \beta \beta_1^2-12 \beta \alpha_1 \omega^3,\\
H_b&=&-12 z \beta \alpha_1^4-24 z \beta \alpha_1^3 \omega+2 z\alpha \alpha_1^3+2 z \beta d^2 \alpha_1^2-12 \alpha_1^2 z \beta \omega^2+\alpha_1^2 t \\&& -8 \alpha_1^2 z \beta \beta_1^2+4 z \alpha \alpha_1^2 \omega+2 \alpha_1 t  \omega+4 z \beta d^2 \alpha_1 \omega+2 \alpha_1 z \alpha \beta_1^2+2 \alpha_1 z \alpha \omega^2\\&&+8 \alpha_1 \omega z \beta \beta_1^2+2 z \beta d^2 \omega^2+4 z \beta  \omega^2 \beta_1^2+z+t \omega^2+4 z \beta \beta_1^4+t \beta_1^2+2 z d^2 \beta \beta_1^2,\end{eqnarray*}

\begin{eqnarray*} \bar A(t,z)&:=&768 d^2 z^2 \alpha \alpha_1^4 \beta \omega^3-16 d^2 z^2 \alpha^2 \alpha_1^6-20 d^4 z^2 \alpha^2 \alpha_1^4+3 \omega^4+96 d^2 z^2 \alpha_1^4 \beta\\&&
+192 d^2 z^2 \alpha \alpha_1^3 \beta \omega^4+3 \alpha_1^4
+384 d^2 t z \alpha_1^3 \beta \omega^3-96 d^6 t z \beta \omega \alpha_1-64 d^2 t z \alpha \alpha_1^2 \omega^3\\&&-16 d^2 t z \alpha \alpha_1 \omega^4
+1152 d^2 z^2 \alpha \alpha_1^5 \beta \omega^2+384 d^2 t z \alpha_1^5 \beta \omega+576 d^2 t z \alpha_1^4 \beta \omega^2\\&&
-192 d^4 t z \beta \omega^2 \alpha_1^2-192 d^4 t z \beta \omega^3 \alpha_1-16 d^2 z^2 \alpha \alpha_1^3+3 \alpha_1^2 d^2-24 d^2 t^2 \omega^2 \alpha_1^2\\&&-16 d^2 t^2 \omega^3 \alpha_1
-16 d^2 t^2 \alpha_1^3 \omega-144 d^4 z^2 \beta \alpha_1^2-48 d^4 z^2 \beta \omega^2+96 d^2 t z \beta \omega^4 \alpha_1^2\\&&
-144 d^8 z^2 \beta^2 \alpha_1^2-144 d^8 z^2 \beta^2 \omega^2-8 d^4 t^2 \omega \alpha_1-8 d^2 t z \alpha_1^2-8 d^2 t z \omega^2\\&&
-64 d^2 t z \alpha \alpha_1^4 \omega-96 d^2 t z \alpha \alpha_1^3 \omega^2-32 d^4 t z \alpha \alpha_1^2 \omega
-16 d^4 t z \alpha \alpha_1 \omega^2+8 i z d^2 \alpha \omega^4\\&&
-96 i z d^2 \beta \alpha_1^5+8 i z d^2 \alpha \alpha_1^4
+8 i d^4 z \alpha \omega^2+8 i d^4 z \alpha \alpha_1^2-96 i d^4 z \alpha_1^3 \beta-4 d^4 t^2 \omega^2\\&&
-8 i z d^2 \alpha_1-8 i z d^2 \omega
-48 d^4 z^2 \alpha^2 \alpha_1^3 \omega-40 d^4 z^2 \alpha^2 \alpha_1^2 \omega^2-2304 d^2 z^2 \alpha_1^7 \beta^2 \omega\\&&
-48 d^6 t z \beta \alpha_1^2-48 d^6 t z \beta \omega^2+8 d^4 z^2 \alpha \omega-4 d^6 z^2 \alpha^2 \omega^2-4 d^6 z^2 \alpha^2 \alpha_1^2
-4 d^4 z^2 \alpha^2 \omega^4+8 d^4 z^2 \alpha \alpha_1\\&&
-864 d^6 z^2 \alpha_1^2 \beta^2 \omega^2-576 d^6 z^2 \alpha_1 \beta^2 \omega^3
-32 d^2 z^2 \alpha \alpha_1^2 \omega-16 d^2 z^2 \alpha \alpha_1 \omega^2-4 d^2 t^2 \alpha_1^4-4 d^2 t^2 \omega^4\\&&
-4 d^4 t^2 \alpha_1^2-1152 d^4 z^2 \alpha_1^5 \beta^2 \omega-576 d^6 z^2 \alpha_1^3 \beta^2 \omega-576 d^4 z^2 \alpha_1^4 \beta^2 \omega^2+6 d^2 \alpha_1 \omega-8 d^6 z^2 \alpha^2 \alpha_1 \omega\\&&
-96 i d^4 z \alpha_1 \beta \omega^2-192 i d^4 z \alpha_1^2 \beta \omega
+16 i d^4 z \alpha \alpha_1 \omega-384 i z d^2 \beta \alpha_1^4 \omega-384 i z d^2 \beta \alpha_1^2 \omega^3\\&&
+32 i z d^2 \alpha \alpha_1^3 \omega
-576 i z d^2 \beta \alpha_1^3 \omega^2+48 i z d^2 \alpha \omega^2 \alpha_1^2+32 i z d^2 \alpha \omega^3 \alpha_1-2304 d^2 z^2 \alpha_1^5 \beta^2 \omega^3\\&&
-3456 d^2 z^2 \alpha_1^6 \beta^2 \omega^2-576 d^2 z^2 \alpha_1^4 \beta^2 \omega^4+3 d^2 \omega^2+192 d^2 z^2 \alpha \alpha_1^7 \beta\\&&
-96 i
 z d^2 \beta \alpha_1 \omega^4+12 \omega^3 \alpha_1+18 \alpha_1^2 \omega^2-64 d^2 z^2 \alpha^2 \alpha_1^5 \omega-48 d^4 t z \beta \omega^4\\&&
-96 d^2 z^2 \alpha^2 \alpha_1^4 \omega^2-64 d^2 z^2 \alpha^2 \alpha_1^3 \omega^3-16 d^2 z^2 \alpha^2 \alpha_1^2 \omega^4+48 d^4 t z \alpha_1^4 \beta\\&&
+192 d^4 z^2 \alpha \alpha_1^5 \beta-16 d^4 z^2 \alpha^2 \alpha_1 \omega^3+96 d^2 t z \alpha_1^6 \beta-16 d^2 t z \alpha \alpha_1^5\\&&
-16 d^4 t z \alpha \alpha_1^3-576 d^2 z^2 \alpha_1^8 \beta^2-576 d^4 z^2 \alpha_1^6 \beta^2-144 d^6 z^2 \beta^2 \omega^4-144 d^6 z^2 \alpha_1^4 \beta^2\\&&
-288 d^8 z^2 \beta^2 \omega \alpha_1+192 d^2 z^2 \alpha_1^3 \beta \omega+96 d^2 z^2 \alpha_1^2 \beta \omega^2+12 \alpha_1^3 \omega-16 d^2 t z \omega \alpha_1\\&&
-192 d^4 z^2 \beta \omega \alpha_1+768 d^2 z^2 \alpha \alpha_1^6 \beta \omega+192 d^4 z^2 \alpha \alpha_1^3 \beta \omega^2
+384 d^4 z^2 \alpha \alpha_1^4 \beta \omega-4 z^2 d^2,
\end{eqnarray*}

\begin{eqnarray*}\bar B(t,z)&:=&-768 d^2 z^2 \alpha \alpha_1^4 \beta \omega^3+16 d^2 z^2 \alpha^2 \alpha_1^6+20 d^4 z^2 \alpha^2 \alpha_1^4+\omega^4-96 d^2 z^2 \alpha_1^4 \beta\\&&
-192 d^2 z^2 \alpha \alpha_1^3 \beta \omega^4+\alpha_1^4-384 d^2 t z \alpha_1^3 \beta \omega^3+96 d^6 t z \beta \omega \alpha_1+64 d^2 t z \alpha \alpha_1^2 \omega^3\\&&
+16 d^2 t z \alpha \alpha_1 \omega^4-1152 d^2 z^2 \alpha \alpha_1^5 \beta \omega^2-384 d^2 t z \alpha_1^5 \beta \omega-576 d^2 t z \alpha_1^4 \beta \omega^2\\&&
+192 d^4 t z \beta \omega^2 \alpha_1^2+192 d^4 t z \beta \omega^3 \alpha_1+16 d^2 z^2 \alpha \alpha_1^3+\alpha_1^2 d^2+24 d^2 t^2 \omega^2 \alpha_1^2\\&&
+16 d^2 t^2 \omega^3 \alpha_1+16 d^2 t^2 \alpha_1^3 \omega+144 d^4 z^2 \beta \alpha_1^2+48 d^4 z^2 \beta \omega^2-96 d^2 t z \beta \omega^4 \alpha_1^2\\&&
+144 d^8 z^2 \beta^2 \alpha_1^2+144 d^8 z^2 \beta^2 \omega^2+8 d^4 t^2 \omega \alpha_1+8 d^2 t z \alpha_1^2+8 d^2 t z \omega^2+64 d^2 t z \alpha \alpha_1^4 \omega\\&&
+96 d^2 t z \alpha \alpha_1^3 \omega^2+32 d^4 t z \alpha \alpha_1^2 \omega+16 d^4 t z \alpha \alpha_1 \omega^2+48 d^4 z^2 \alpha^2 \alpha_1^3 \omega\\&&
+40 d^4 z^2 \alpha^2 \alpha_1^2 \omega^2+2304 d^2 z^2 \alpha_1^7 \beta^2 \omega+48 d^6 t z \beta \alpha_1^2+48 d^6 t z \beta \omega^2-8 d^4 z^2 \alpha \omega\\&&
+4 d^6 z^2 \alpha^2 \omega^2+4 d^6 z^2 \alpha^2 \alpha_1^2+4 d^4 z^2 \alpha^2 \omega^4-8 d^4 z^2 \alpha \alpha_1+864 d^6 z^2 \alpha_1^2 \beta^2 \omega^2\\&&
+576 d^6 z^2 \alpha_1 \beta^2 \omega^3+32 d^2 z^2 \alpha \alpha_1^2 \omega+16 d^2 z^2 \alpha \alpha_1 \omega^2+4 d^2 t^2 \alpha_1^4+4 d^2 t^2 \omega^4\\&&
+4 d^4 t^2 \omega^2+4 d^4 t^2 \alpha_1^2+1152 d^4 z^2 \alpha_1^5 \beta^2 \omega+576 d^6 z^2 \alpha_1^3 \beta^2 \omega+576 d^4 z^2 \alpha_1^4 \beta^2 \omega^2\\&&
+2 d^2 \alpha_1 \omega+8 d^6 z^2 \alpha^2 \alpha_1 \omega+2304 d^2 z^2 \alpha_1^5 \beta^2 \omega^3+3456 d^2 z^2 \alpha_1^6 \beta^2 \omega^2\\&&
+576 d^2 z^2 \alpha_1^4 \beta^2 \omega^4+d^2 \omega^2-192 d^2 z^2 \alpha \alpha_1^7 \beta+4 \omega^3 \alpha_1+6 \alpha_1^2 \omega^2\\&&
+64 d^2 z^2 \alpha^2 \alpha_1^5 \omega+48 d^4 t z \beta \omega^4+96 d^2 z^2 \alpha^2 \alpha_1^4 \omega^2+64 d^2 z^2 \alpha^2 \alpha_1^3 \omega^3\\&&
+16 d^2 z^2 \alpha^2 \alpha_1^2 \omega^4-48 d^4 t z \alpha_1^4 \beta-192 d^4 z^2 \alpha \alpha_1^5 \beta+16 d^4 z^2 \alpha^2 \alpha_1 \omega^3\\&&
-96 d^2 t z \alpha_1^6 \beta+16 d^2 t z \alpha \alpha_1^5+16 d^4 t z \alpha \alpha_1^3+576 d^2 z^2 \alpha_1^8 \beta^2+576 d^4 z^2 \alpha_1^6 \beta^2\\&&
+144 d^6 z^2 \beta^2 \omega^4+144 d^6 z^2 \alpha_1^4 \beta^2+288 d^8 z^2 \beta^2 \omega \alpha_1-192 d^2 z^2 \alpha_1^3 \beta \omega\\&&
-96 d^2 z^2 \alpha_1^2 \beta \omega^2+4 \alpha_1^3 \omega+16 d^2 t z \omega \alpha_1+192 d^4 z^2 \beta \omega \alpha_1\\&&
-768 d^2 z^2 \alpha \alpha_1^6 \beta \omega-192 d^4 z^2 \alpha \alpha_1^3 \beta \omega^2-384 d^4 z^2 \alpha \alpha_1^4 \beta \omega+4 z^2 d^2.
\end{eqnarray*}
\end{widetext}


\begin{widetext}

\begin{figure}[h!]
\centering
\raisebox{0.85in}{($|E|^2$)}\includegraphics[scale=0.18]{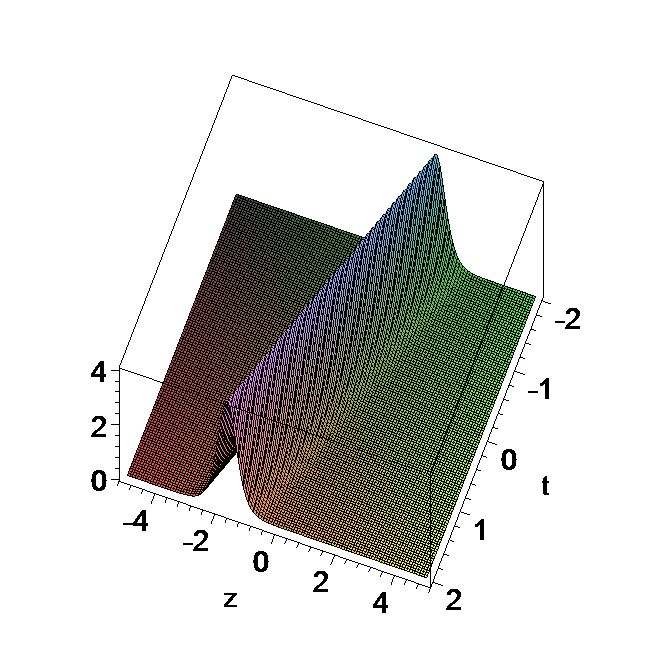}
\hskip 0.03cm
\raisebox{0.85in}{($|p|^2$)}\raisebox{-0.1cm}{\includegraphics[scale=0.18]{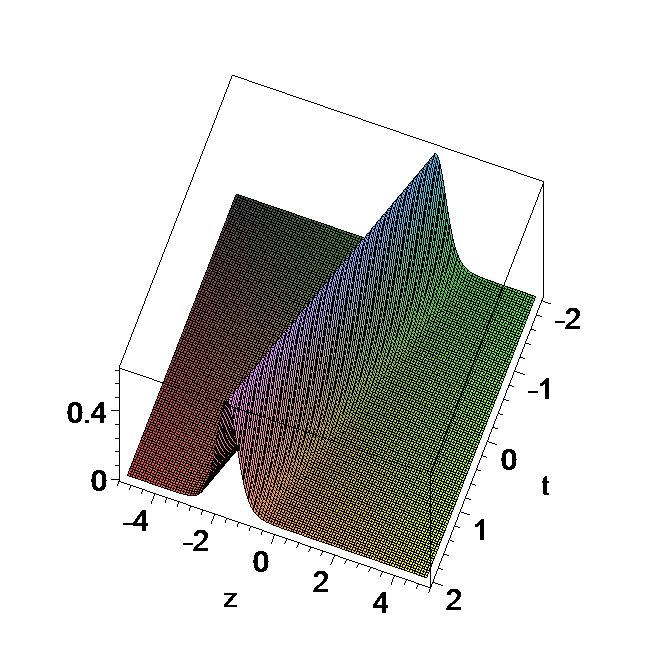}}
\hskip 0.03cm
\raisebox{0.85in}{($\eta$)}\raisebox{-0.1cm}{\includegraphics[scale=0.18]{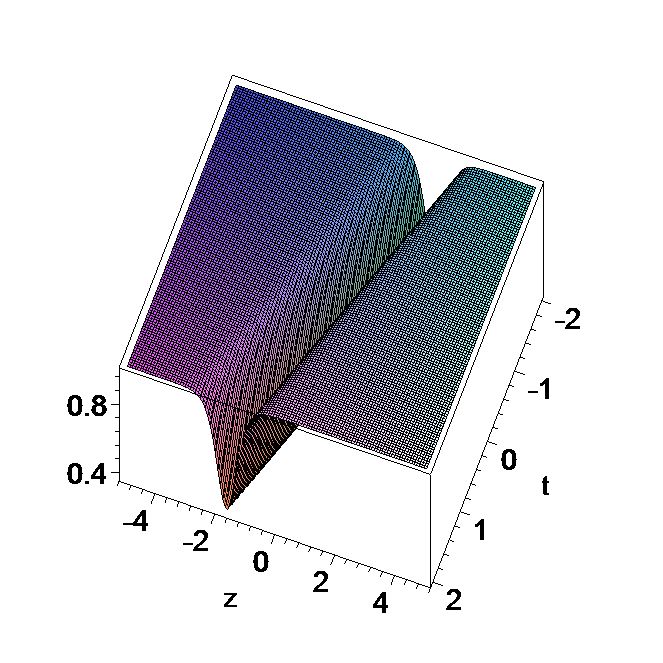}}
 \caption{\small (color online) One solition solution $(E, p,\eta)$  of the H-MB system with $\omega=1.5,\alpha_1=0.5,\beta_1=1,\alpha=2,\beta=-1$.}\label{1solitonH-MB}
\end{figure}

\begin{figure}[h!]
\centering
\raisebox{0.85in}{($|E|^2$)}\includegraphics[scale=0.20]{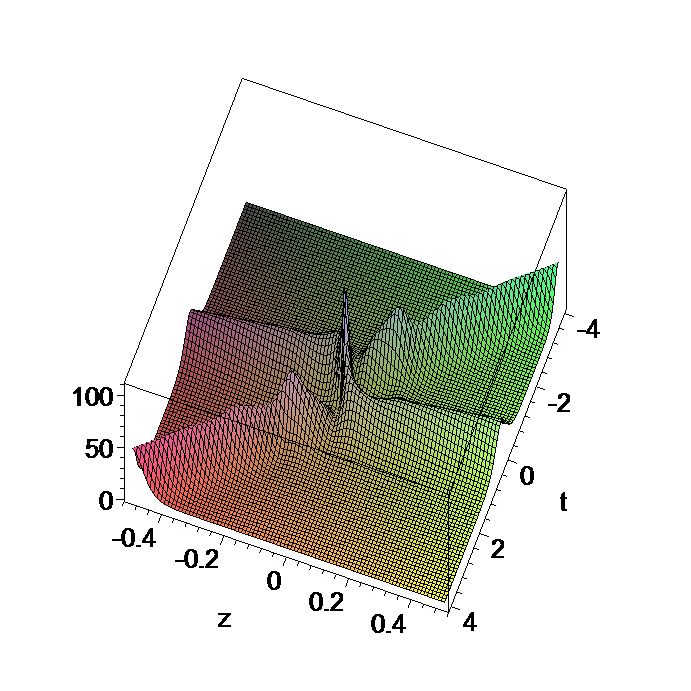}
\hskip 0.03cm
\raisebox{0.85in}{($|p|^2$)}\raisebox{-0.1cm}{\includegraphics[scale=0.20]{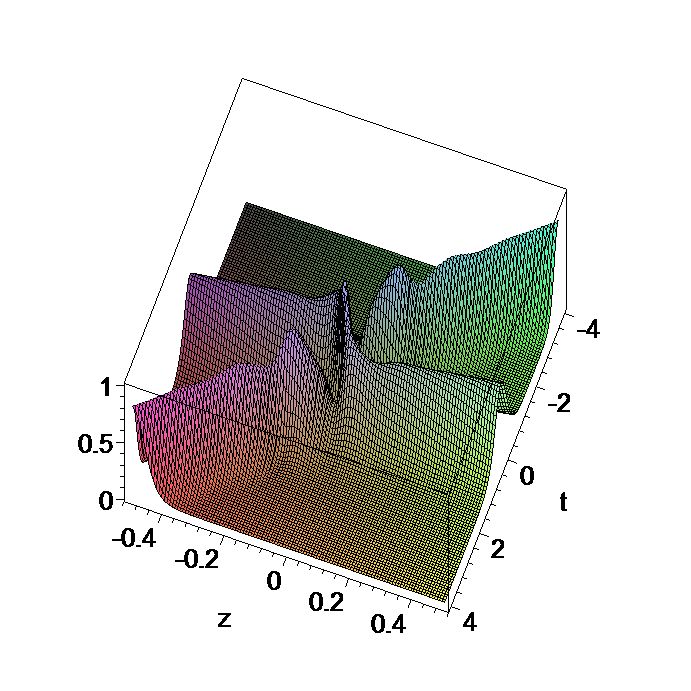}}
\hskip 0.03cm
\raisebox{0.85in}{($\eta$)}\raisebox{-0.1cm}{\includegraphics[scale=0.20]{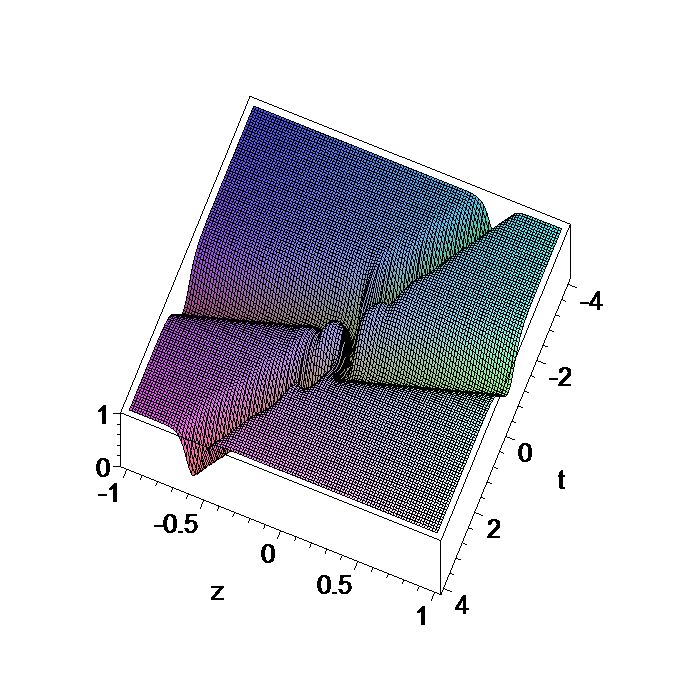}}
\caption{\small (color online) Two solition solution $(E, p,\eta)$  of the H-MB system with $\omega=1.5,\alpha_1=0.5,\beta_1=1,\alpha=2, \beta=-1$.}\label{2solitonH-MB}
\end{figure}

\begin{figure}[h!]
\centering
\raisebox{0.85in}{($|E|^2$)}\includegraphics[scale=0.19]{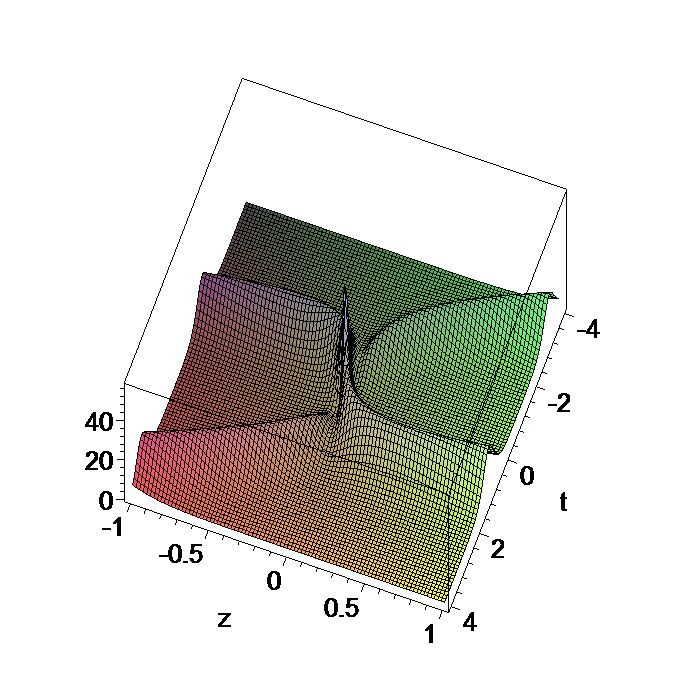}
\hskip 0.03cm
\raisebox{0.85in}{($|p|^2$)}\raisebox{-0.1cm}{\includegraphics[scale=0.19]{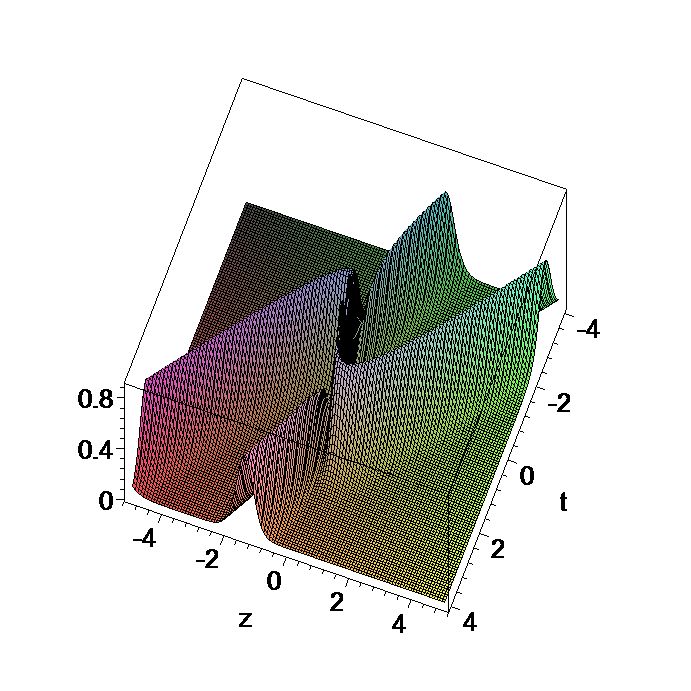}}
\hskip 0.03cm
\raisebox{0.85in}{($\eta$)}\raisebox{-0.1cm}{\includegraphics[scale=0.19]{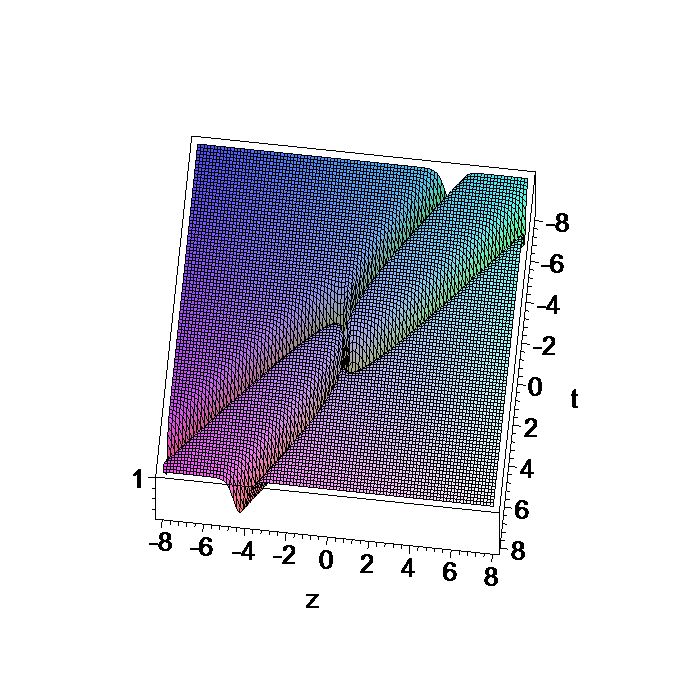}}
 \caption{\small (color online) One positon solution $(E,p,\eta)$  of the H-MB system when $\omega=1.5,\alpha_1=0.5,\beta_1=1,\alpha=2, \beta=-1$.}\label{Hpositon}
\end{figure}

\begin{figure}[h!]
\centering
\raisebox{0.85in}{($|E|^2$)}\includegraphics[scale=0.24]{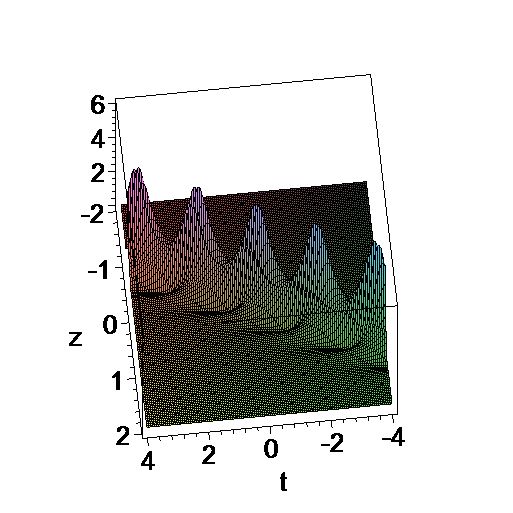}
\hskip 0.03cm
\raisebox{0.85in}{($|p|^2$)}\raisebox{-0.1cm}{\includegraphics[scale=0.18]{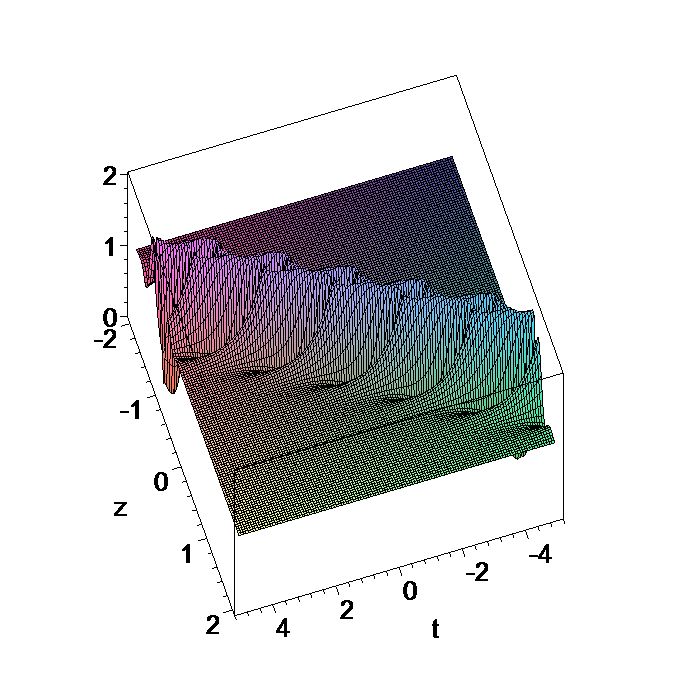}}
\hskip 0.03cm
\raisebox{0.85in}{($\eta$)}\raisebox{-0.1cm}{\includegraphics[scale=0.18]{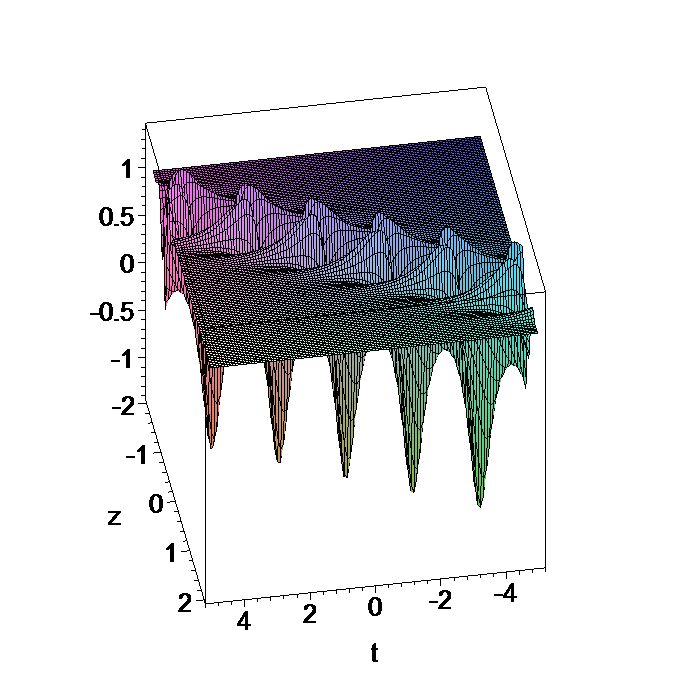}}
 \caption{\small (color online) Breather solution $(E,p,\eta)$  of the H-MB  equation when $\omega=0.5,d=0.5,n_1=1,\alpha_1=-1,\beta_1=1,\alpha=2, \beta=-1$.} \label{Hbreather}
\end{figure}

\begin{figure}[h!]
\centering
\raisebox{0.85in}{($|E|^2$)}\includegraphics[scale=0.24]{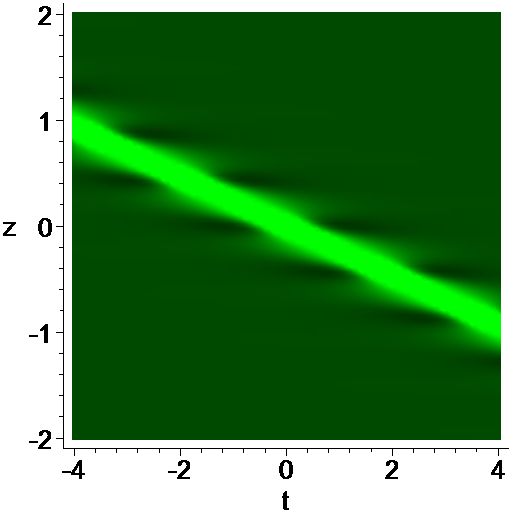}
\hskip 0.03cm
\raisebox{0.85in}{($|p|^2$)}\raisebox{-0.1cm}{\includegraphics[scale=0.18]{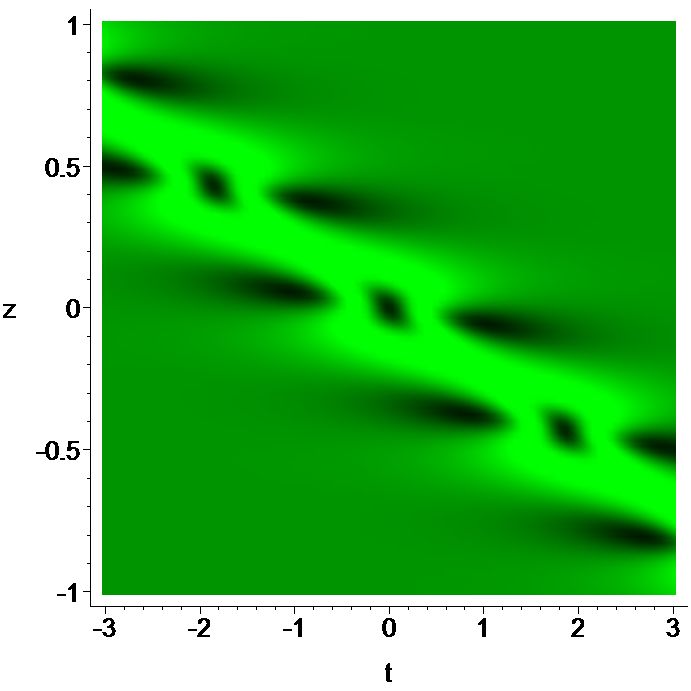}}
\hskip 0.03cm
\raisebox{0.85in}{($\eta$)}\raisebox{-0.1cm}{\includegraphics[scale=0.18]{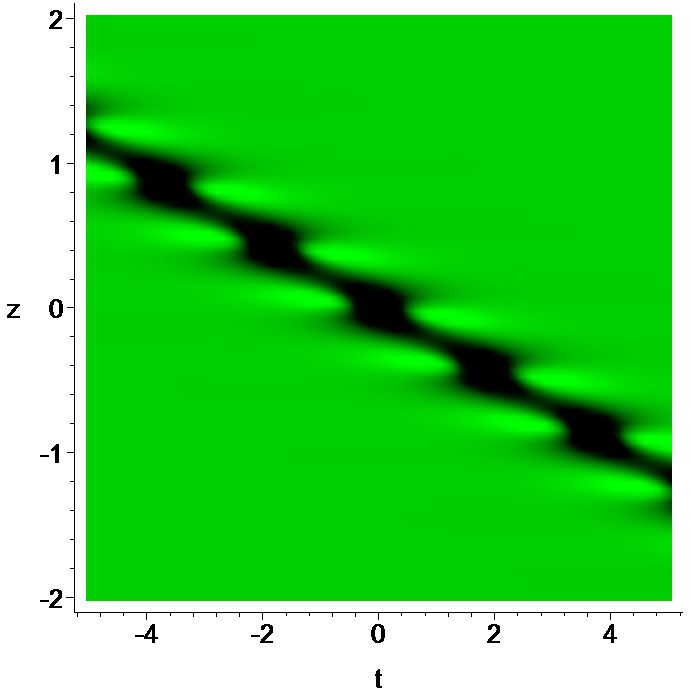}}
 \caption{\small (color online) Breather solution $(E,p,\eta)$  of the H-MB equation  when $\omega=0.5,d=0.5,\alpha_1=-1,\beta_1=1,\alpha=2, \beta=-1$.} \label{Hbreatherdensity}
\end{figure}

\begin{figure}[h!]
\centering
\raisebox{0.85in}{($|E|^2$)}\includegraphics[scale=0.26]{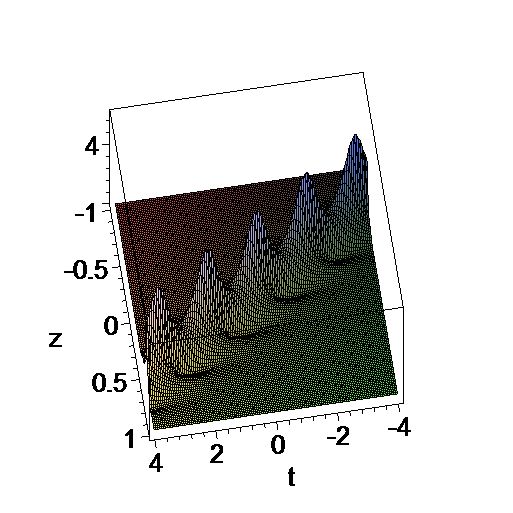}
\hskip 0.03cm
\raisebox{0.85in}{($|p|^2$)}\raisebox{-0.1cm}{\includegraphics[scale=0.20]{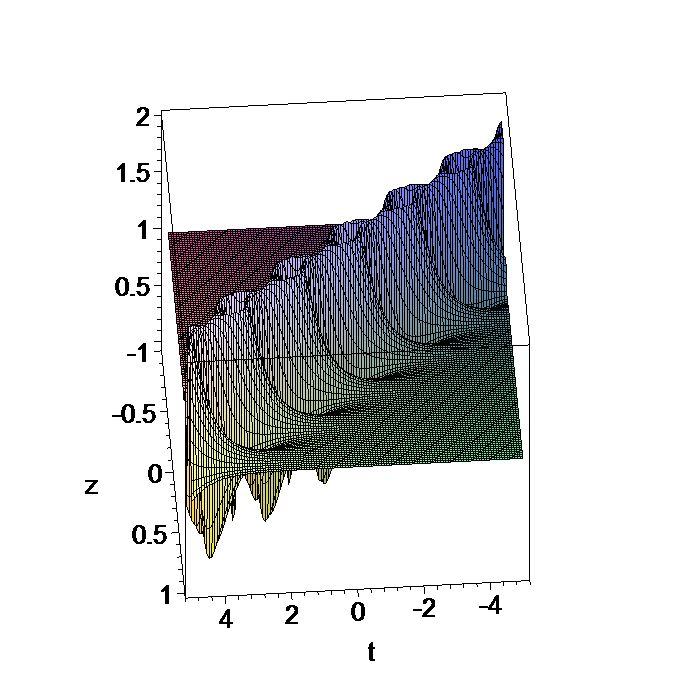}}
\hskip 0.03cm
\raisebox{0.85in}{($\eta$)}\raisebox{-0.1cm}{\includegraphics[scale=0.18]{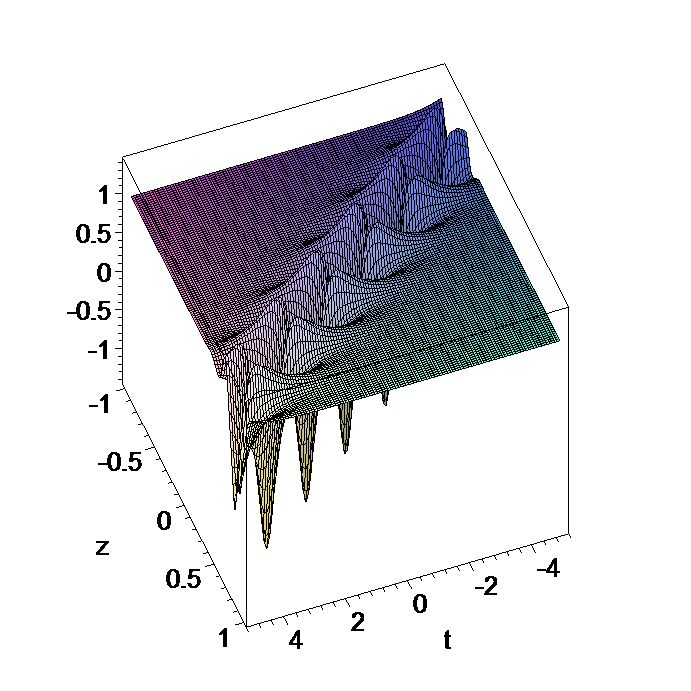}}
 \caption{\small (color online) Breather solution $(E,p,\eta)$  of the (0,1)-CMKdV-MB  equation  when $\omega=0.5,d=0.5,\alpha_1=-1,\beta_1=1,\alpha=0, \beta=1$.} \label{01breather}
\end{figure}

\begin{figure}[h!]
\centering
\raisebox{0.25in}{(2,-1)}\includegraphics[scale=0.18]{H-MB-breathery-real-density}
\hskip 0.03cm
\raisebox{0.25in}{(1,0)}\raisebox{-0.1cm}{\includegraphics[scale=0.18]{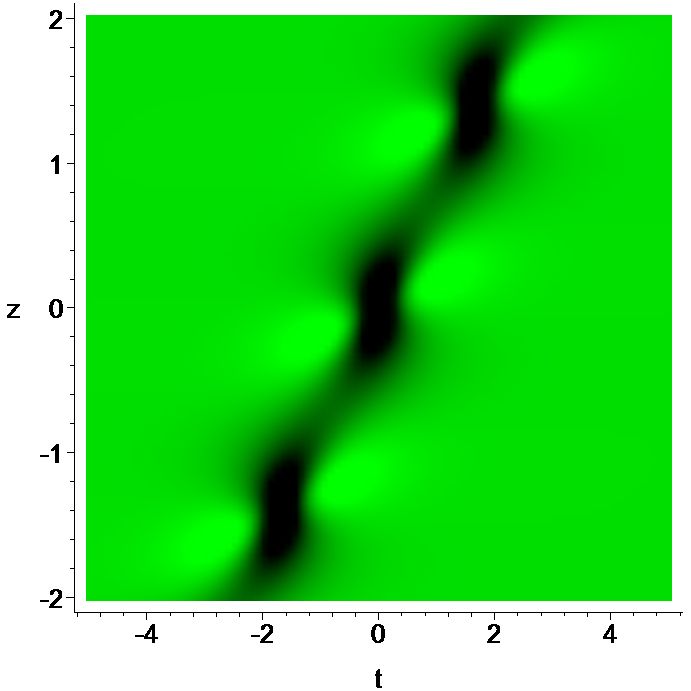}}
\hskip 0.03cm
\raisebox{0.25in}{(0,1)}\raisebox{-0.1cm}{\includegraphics[scale=0.18]{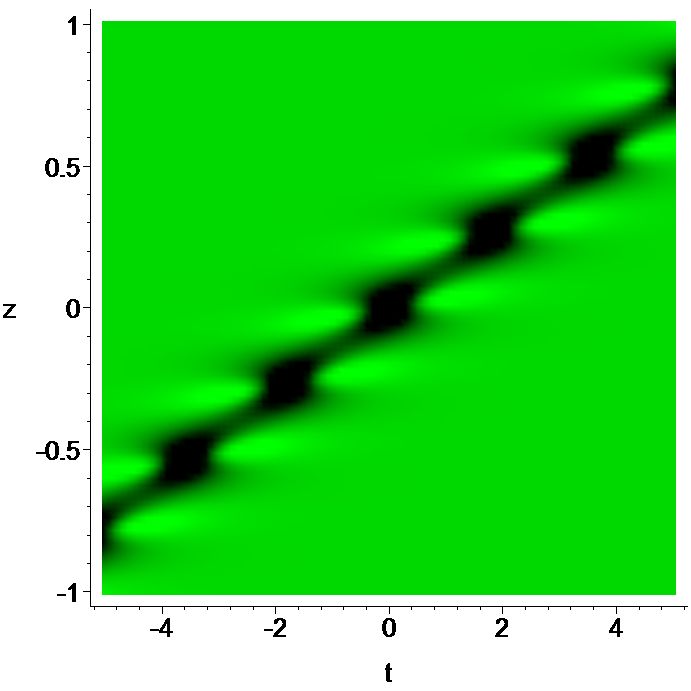}}
\hskip 0.03cm
\raisebox{0.25in}{(1,1)}\raisebox{-0.1cm}{\includegraphics[scale=0.18]{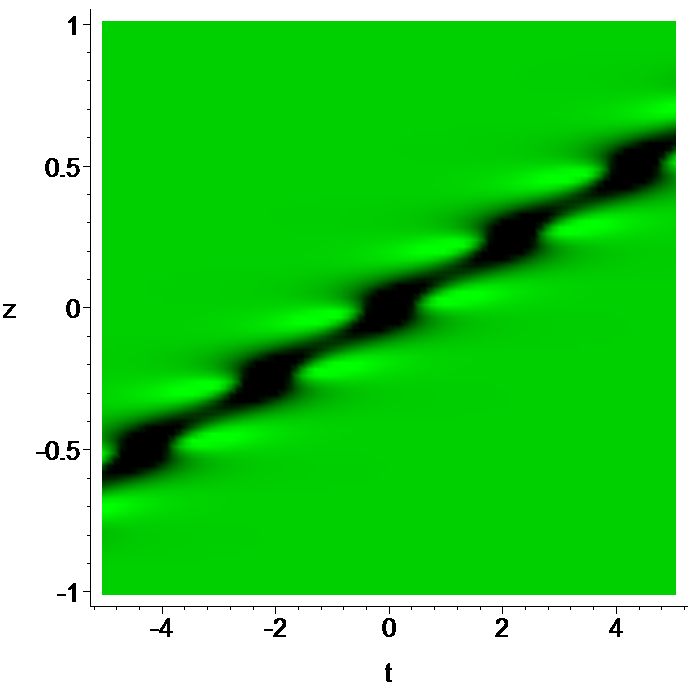}}
\hskip 0.03cm
\raisebox{0.25in}{(-1,0)}\raisebox{-0.1cm}{\includegraphics[scale=0.18]{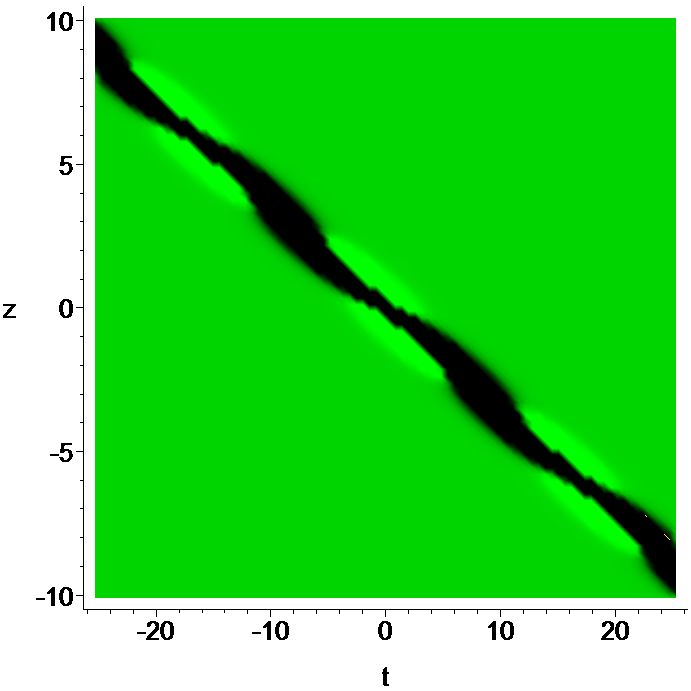}}
\hskip 0.03cm
\raisebox{0.25in}{(0,-1)}\raisebox{-0.1cm}{\includegraphics[scale=0.18]{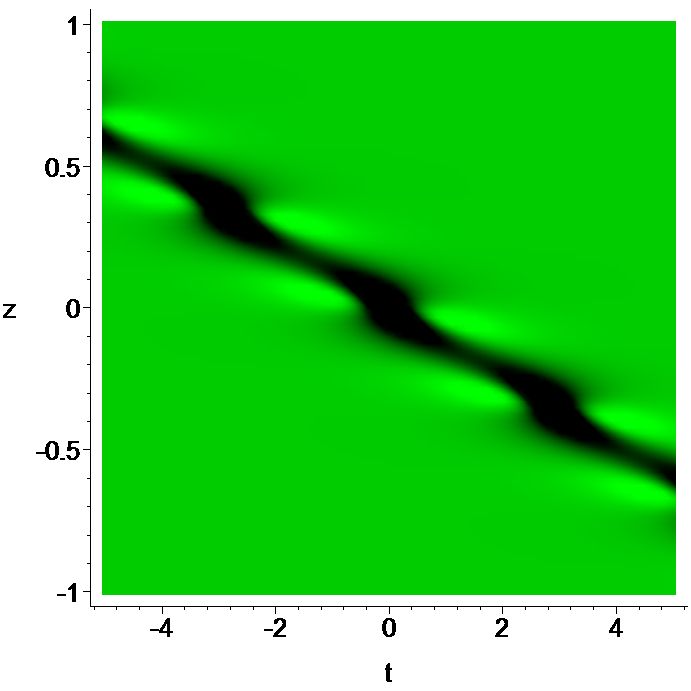}}
 \caption{\small (color online) Breather solution $\eta$ for different values of $(\alpha, \beta)$ of the H-MB system when $\omega=0.5,d=0.5,\alpha_1=-1,\beta_1=1.$} \label{breatherdirection}
\end{figure}

\begin{figure}[h!]
\centering
\raisebox{0.85in}{($|E|^2$)}\includegraphics[scale=0.22]{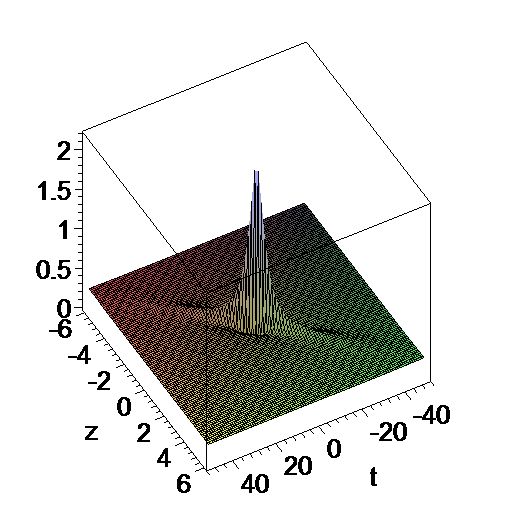}
\hskip 0.03cm
\raisebox{0.85in}{($|p|^2$)}\raisebox{-0.1cm}{\includegraphics[scale=0.18]{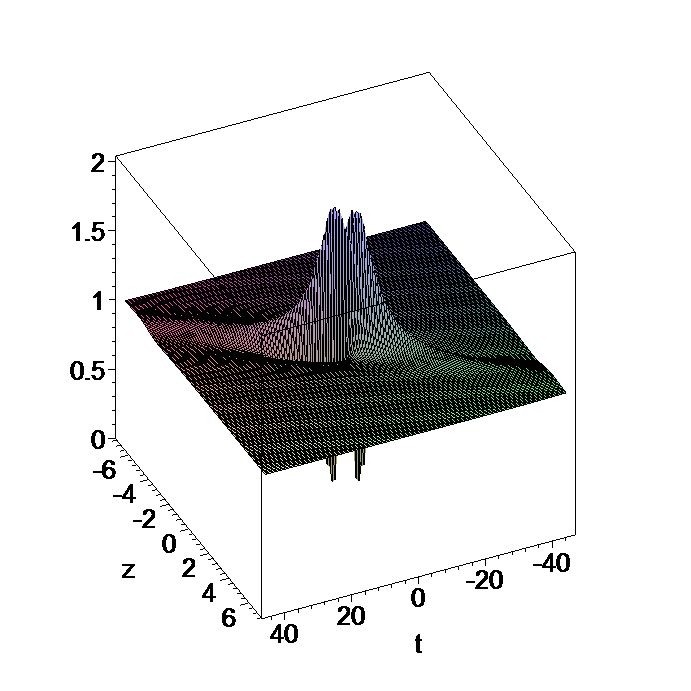}}
\hskip 0.03cm
\raisebox{0.85in}{($\eta$)}\raisebox{-0.1cm}{\includegraphics[scale=0.18]{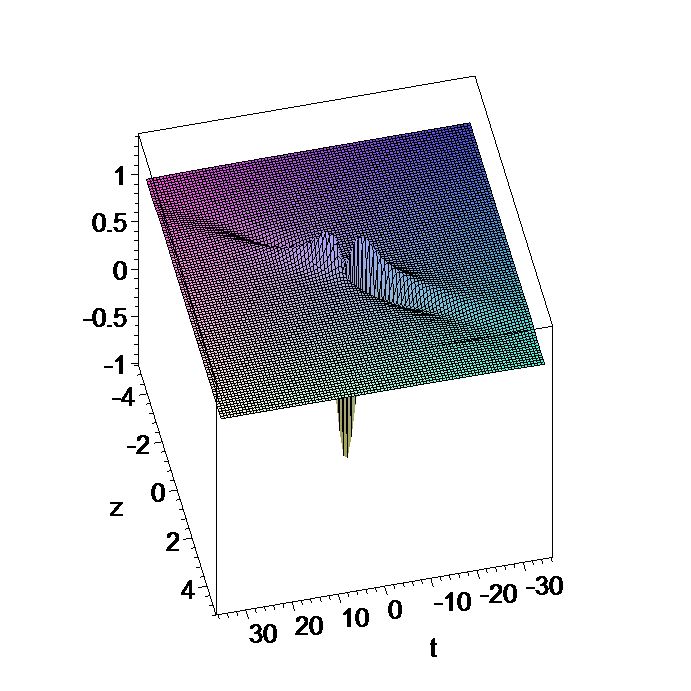}}
 \caption{\small (color online) Rogue wave  solution $(E,p,\eta)$  of the H-MB  system when $\omega=0.5,d=0.5,\alpha_1=-1,\beta_1=1,\alpha=2, \beta=-1$.} \label{H-MB-rogue}
\end{figure}

\begin{figure}[h!]
\centering
\raisebox{0.85in}{($|E|^2$)}\includegraphics[scale=0.25]{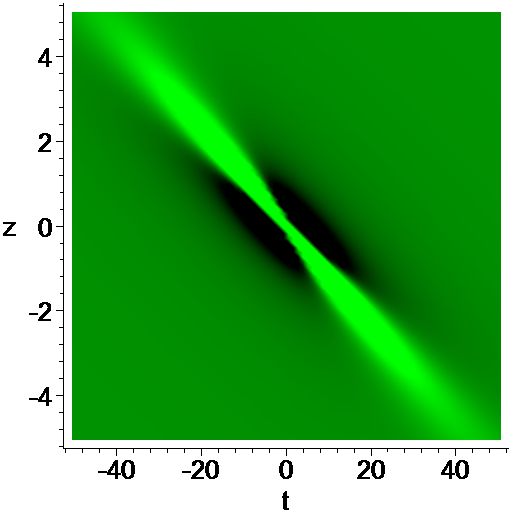}
\hskip 0.03cm
\raisebox{0.85in}{($|p|^2$)}\raisebox{-0.1cm}{\includegraphics[scale=0.19]{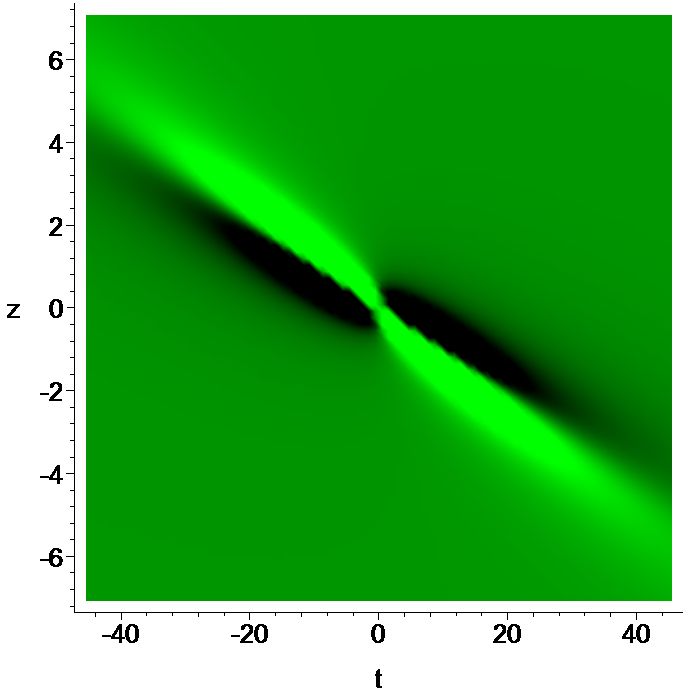}}
\hskip 0.03cm
\raisebox{0.85in}{($\eta$)}\raisebox{-0.1cm}{\includegraphics[scale=0.31]{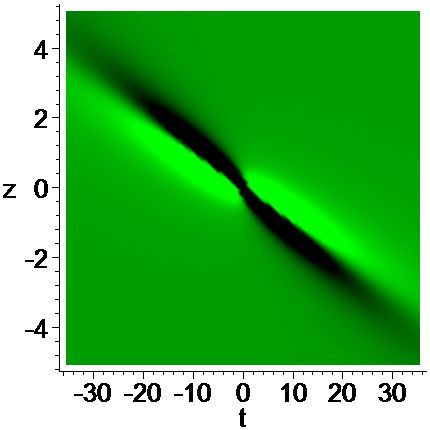}}
 \caption{\small (color online) Rogue wave solution $(E,p,\eta)$  of the H-MB  equation when $\omega=0.5,d=0.5,\alpha_1=-1,\beta_1=1,\alpha=2, \beta=-1$.} \label{H-MB-rogue-density}
\end{figure}

\begin{figure}[h!]
\centering
\raisebox{0.85in}{($|p|^2$)}\raisebox{-0.1cm}{\includegraphics[scale=0.18]{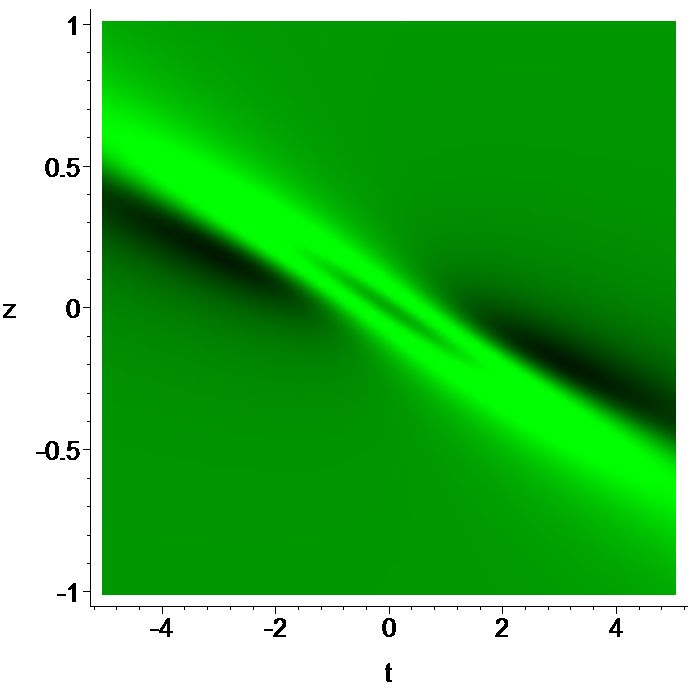}}
 \caption{\small (color online) Enlarged graph of rogue wave  solution $p$ (in Fig. \ref{H-MB-rogue-density}) of the H-MB  system when $\omega=0.5,d=0.5,\alpha_1=-1,\beta_1=1,\alpha=2, \beta=-1$.} \label{H-MB-rogue-density-close}
\end{figure}

\begin{figure}[h!]
\centering
\raisebox{0.85in}{($|E|^2$)}\includegraphics[scale=0.25]{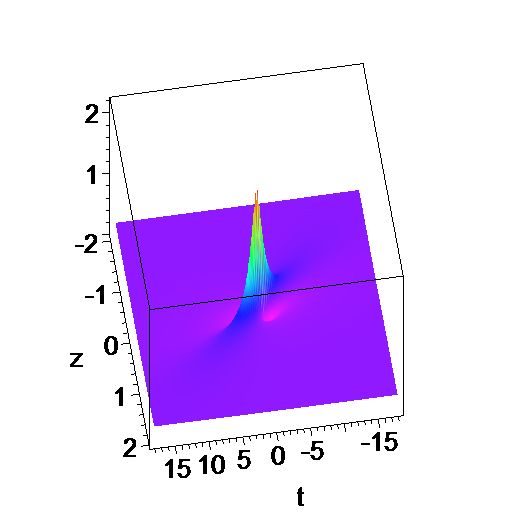}
\hskip 0.03cm
\raisebox{0.85in}{($|p|^2$)}\raisebox{-0.1cm}{\includegraphics[scale=0.19]{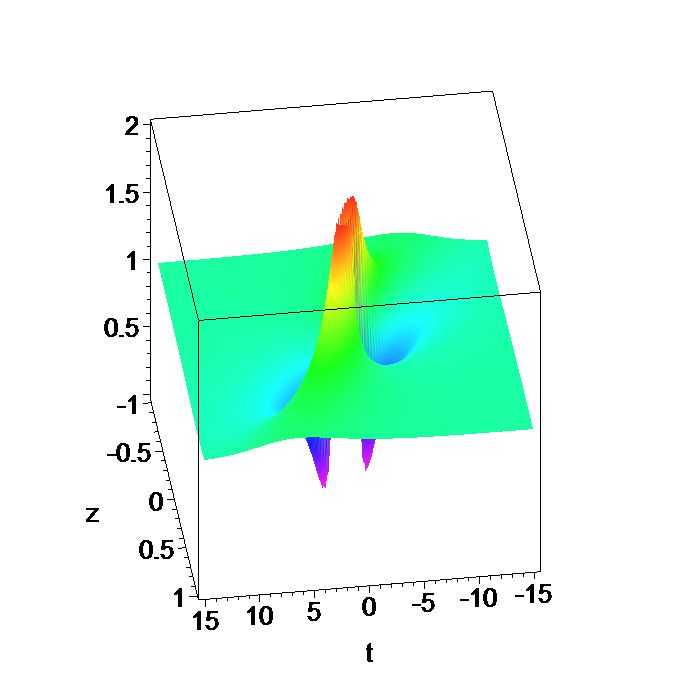}}
\hskip 0.03cm
\raisebox{0.85in}{($\eta$)}\raisebox{-0.1cm}{\includegraphics[scale=0.19]{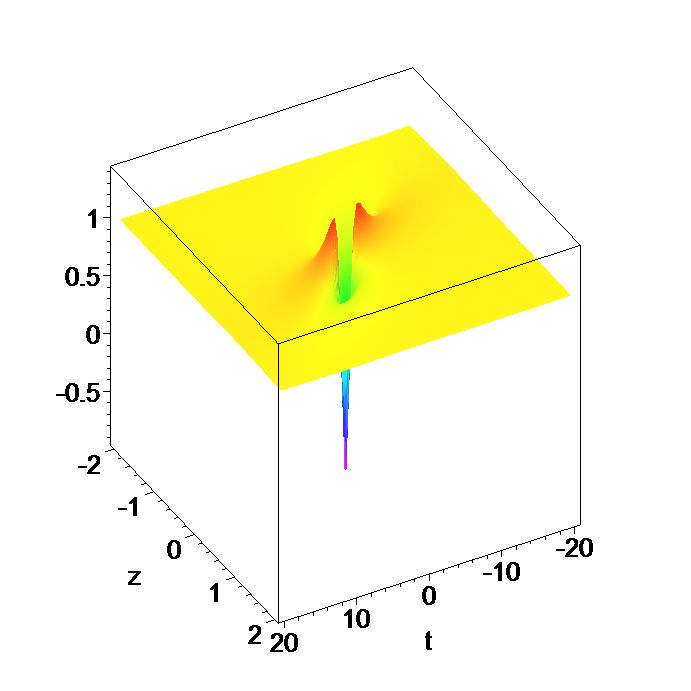}}
 \caption{\small (color online) Rogue wave solution $(E,p,\eta)$  of the (0,1)-CMKdV-MB  equation when $\omega=0.5,d=0.5,\alpha_1=-1,\beta_1=1,\alpha=0, \beta=1$.} \label{01-MB-rogue}
\end{figure}

\begin{figure}[h!]
\centering
\raisebox{0.85in}{($|E|^2$)}\includegraphics[scale=0.28]{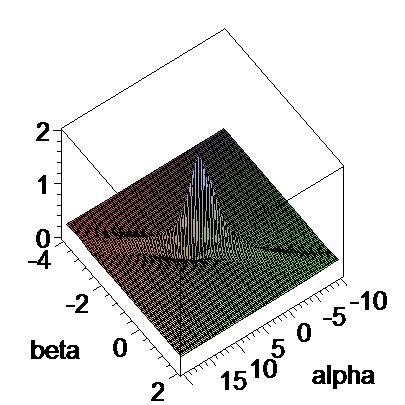}
\hskip 0.03cm
\raisebox{0.85in}{($|p|^2$)}\raisebox{-0.1cm}{\includegraphics[scale=0.19]{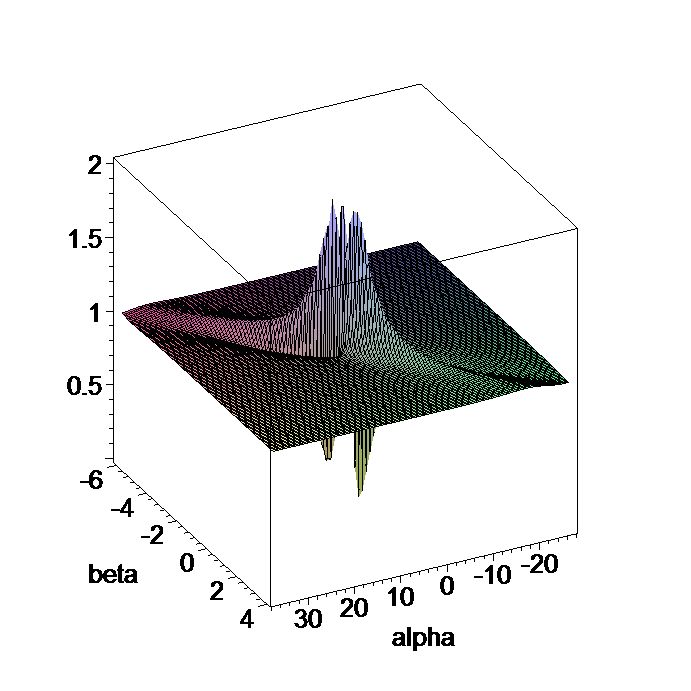}}
\hskip 0.03cm
\raisebox{0.85in}{($\eta$)}\raisebox{-0.1cm}{\includegraphics[scale=0.19]{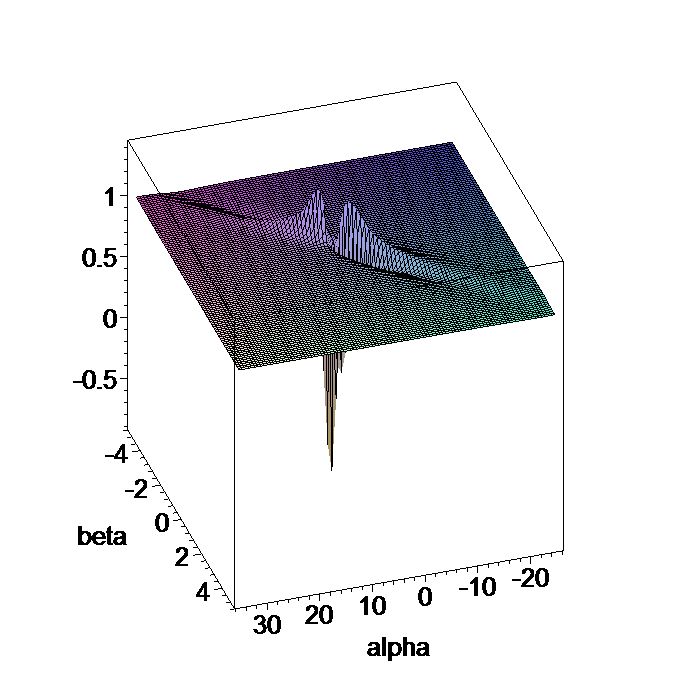}}
 \caption{\small (color online) Rogue wave  solution $(E,p,\eta)$  of the H-MB  equation when $\omega=0.5,d=0.5,\alpha_1=-1,\beta_1=1,t=1,z=1$.} \label{t=1z=1alphabeta-rogue}
\end{figure}

\end{widetext}



\end{document}